\documentclass[a4paper,11pt]{article}
\pdfoutput=1 

\usepackage{jheppub} 

\usepackage[T1]{fontenc} 

\usepackage{amsmath}

\newcommand{\be}{\begin{equation}}
\newcommand{\ee}{\end{equation}}
\newcommand{\bea}{\begin{eqnarray}}
\newcommand{\eea}{\end{eqnarray}}
\newcommand{\nn}{\nonumber}
\newcommand{\cC}{\mathcal{C}}

\newcommand{\cN}{\mathcal{N}}

\newcommand{\cS}{\mathcal{S}}

\newcommand{\bZ}{\mathbb{Z}}

\newcommand{\fc}{\mathfrak{c}}
\newcommand{\fp}{\mathfrak{p}}

\newcommand{\StarPt}{\mathord{\includegraphics[height=2.7ex]{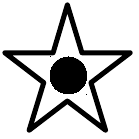}}}
\newcommand{\StarPts}{\mathord{\includegraphics[height=1.4ex]{StarPt}}}

\definecolor{darkgreen}{rgb}{0,0.3,0}

\definecolor{usuzumiiro}{cmyk}{0.08,0.05,0.06,0.55}

\preprint{DESY-15-248, KIAS-P15064,
RIKEN-STAMP-22}

\title{\boldmath Spectral curves of $\cN=1$ theories of class $\cS_k$}

\author[a,b,]{Ioana Coman,}
\author[a,c]{Elli Pomoni,}
\author[d]{Masato Taki}
\author[e]{and Futoshi Yagi}

\affiliation[a]{DESY Theory Group, Notkestra{\ss}e 85, 22607 Hamburg, Germany}
\affiliation[b]{Department of Theoretical Physics, NIPNE,
Str. Reactorului 30, 077125, Magurele, Romania}
\affiliation[c]{Physics Division, National Technical University of Athens,
15780 Zografou Campus, Athens, Greece}
\affiliation[d]{iTHES Research Group and Mathematical Physics Laboratory, RIKEN Nishina Center, Saitama 351-0198, Japan}
\affiliation[e]{Korea Institute for Advanced Study (KIAS)
85 Hoegiro Dongdaemun-gu, 130-722, Seoul, Korea}

\emailAdd{ioana.coman@desy.de}
\emailAdd{elli.pomoni@desy.de}
\emailAdd{taki@riken.jp}
\emailAdd{fyagi@kias.re.kr}

\abstract{

\bigskip

We study the Coulomb branch of  class  $\mathcal{S}_k$  $\mathcal{N} = 1$ SCFTs by constructing and analyzing their spectral curves.

\bigskip

}

\begin{document}
\maketitle
\flushbottom

\section{Introduction}
\label{sec:intro}

A very exiting development in the last 20 years has been the achievement of exact results in $\mathcal{N}=2$ gauge theories in four dimensions. See \cite{Teschner:2014oja,Gaiotto:2014bja} for a recent review.
Starting with the groundbreaking work of Seiberg and Witten \cite{Seiberg:1994rs,Seiberg:1994aj}, who solved 4D  $\mathcal{N}=2$ gauge theories in the IR, and
continuing with the microscopic derivation of  Nekrasov \cite{Nekrasov:2002qd}, today we can compute several observables in the UV thanks to the work of Pestun \cite{Pestun:2007rz} who was able to calculate using Localization techniques the partition function of  $\mathcal{N}=2$ gauge theories on $\mathbb{S}^4$.

In the seminal paper  \cite{Gaiotto:2009we},  Gaiotto  obtained the  so called class $\mathcal{S}$ of $\mathcal{N}=2$ SCFTs $\mathcal{T}_{g,n}$ \cite{Gaiotto:2009we,Gaiotto:2009hg}
 by compactifying (a twisted version of) the
6D $(2, 0)$  SCFT on a Riemann surface $\mathcal{C}_{g,n}$, of genus $g$ and $n$ punctures. For these
SCFTs, the parameter space of exactly marginal gauge couplings  is identified with the moduli space of complex
structures 
of the Riemann surface modulo 
the group  of generalized S-duality
transformations of the 4D theory. This development led to the discovery of the AGT correspondence \cite{Alday:2009aq,Wyllard:2009hg},
 where the partition functions on $\mathbb{S}^4$ of the $\mathcal{T}_{g,n}$ theories are equal to Liouville/Toda field theory correlators on  the corresponding Riemann surface $\mathcal{C}_{g,n}$.
Finally,  in \cite{Gadde:2009kb,Gadde:2011ik,Gadde:2011uv}  another striking 4D/2D duality was discovered.
 The
superconformal index of the 4D theory  associated with the Riemann surface  can be reinterpreted as
a correlation function in a 2D topological QFT living on the Riemann surface. See \cite{Rastelli:2014jja} for a recent review. Invariance of the index
under generalized S-duality transformations translates into associativity of the operator algebra of the 2D TQFT.

A very important long term quest is to understand whether it is possible to make similar progress  with $\mathcal{N} = 1$ gauge theories
which have a far more rich and interesting phase structure \cite{Intriligator:1994sm,Intriligator:1995au}
and also to search for possible relations they may have with 2D CFTs and integrable models.  The class $\mathcal{S}_k$ of $\mathcal{N} = 1$ SCFTs recently proposed by  Gaiotto and Razamat \cite{Gaiotto:2015usa} provides a very promising starting point.
The theories in class  $\mathcal{S}_k$  are obtained via compactification of the 6D $(1,0)$ SCFTs labeled by $N,k$ on a Riemann surface with punctures,
where these theories are obtained from the 6D $(2,0)$ SCFT by orbifolding.
Thus class $\cS_k$ theories can be understood as an orbifolded version of class $\cS$ theories.
Most importantly, in  \cite{Gaiotto:2015usa} the authors were able to recast  the index for class $\mathcal{S}_k$ as a 2D TFT.
 See \cite{Franco:2015jna,Hanany:2015pfa} for works  in  similar spirit and generalizations.

Some of the powerful tools that led to the $\mathcal{N} = 2$  success story include the Seiberg-Witten (spectral) curve, which is an auxiliary curve that contains all the information about the low energy effective theory, and
 the fact that $\mathcal{N} = 2$  theories have string/M-theory constructions (realizations).
 In particular, Witten's M-theory approach \cite{Witten:1997sc} provides a way to realize the SW curve geometrically as an M5-brane configuration and a simple way to obtain it.

Already in \cite{Intriligator:1994sm}, Intriligator and Seiberg pointed out that the SW curve techniques can be applied to the Coulomb branches of  $\mathcal{N}=1$ theories as well.
$\mathcal{N} = 1$ holomorphicity arguments constrain the
form of the superpotential  in the same way that $\mathcal{N} = 2$ holomorphy constrains the
form of the prepotential \cite{Intriligator:1995au}.
 In the Coulomb phase of  $\mathcal{N}=1$ gauge theories there are massless photons in the low-energy theory, whose couplings to the matter fields are described by the
usual gauge kinetic term $\tau_{ij}W^{i\,\alpha} W^j_{\alpha}$,
with $\tau_{ij}$ the effective gauge couplings which are holomorphic
functions of the matter fields. In the $\mathcal{N}=1$ case the determination of the  $\tau_{ij}$
does not imply a complete solution of the theory, but still provides very important information about the low energy theory.
Witten's M-theory approach to the  SW curve \cite{Witten:1997sc} has already been generalized and used to study $\mathcal{N}=1$ theories
\cite{Witten:1997ep,Hori:1997ab}. See also \cite{Bah:2012dg,Xie:2013gma} for a more recent construction of
 a large set of $\mathcal{N}=1$ SCFTs.  Finally,
much progress has been recently made for  $\mathcal{N} = 1$ spectral curves and their relations to generalized Hitchin systems \cite{Bonelli:2013pva,Xie:2013rsa,Xie:2014yya}.

In this paper, inspired by the work of  \cite{Gaiotto:2015usa} we wish to understand the 4D/2D interplay from the point of view of the SW curves. This is how it was
originally  discovered  for the $\mathcal{N}=2$  class $\mathcal{S}$  in \cite{Gaiotto:2009we}.
 We compute the spectral curves for theories in class $\mathcal{S}_k$  and  generalize Gaiotto's construction for the $\mathcal{N}=2$  theories of class $\mathcal{S}$   \cite{Gaiotto:2009we}  by orbifolding.
 We then study how the spectral curves decompose in different S-duality frames. An important object of interest is the type of punctures. For $\mathcal{N}=2$ theories in  class $\mathcal{S}$ we have punctures labelled by Young diagrams, including
 minimal and maximal punctures that correspond to simple poles with symmetry $U(1)$ and $SU(N)$ respectively.
 As we will discover in the sections to come,  for class $\mathcal{S}_k$ the minimal punctures do not correspond to simple poles,  but to branch points.
We find that the spectral curves have a novel $k$-cut structure\footnote{It is not the first time that an extra  $\mathbb{Z}_k$ symmetry
creates cuts  on the SW curve. Already for $\mathcal{N}=2$ theories cuts appear as a consequence of outer-automorphism $\mathbb{Z}_s$ symmetry of the Dynkin diagram \cite{Tachikawa:2010vg} to which the theory corresponds. See also \cite{Tachikawa:2009rb,Nanopoulos:2009xe,Chacaltana:2012ch,Chacaltana:2013oka,Chacaltana:2014nya,Chacaltana:2015bna}.}.

The paper is organized as follows. We begin with a short review of class $\mathcal{S}_k$ in section \ref{review}. In section \ref{4puncturedS} we show how to orbifold the SW curves of theories from class $\mathcal{S}$ with a Lagrangian description. We study the four-punctured sphere, the maximal and minimal punctures. Most importantly we discuss the novel cut structure imposed by the orbifold. In section \ref{sec:Trinion} we take the weak coupling limit and discover the free trinion theory, the curve for the free orbifloded hypermultiplets.  Next in section \ref{Closing}, we begin with the curve of the
four-punctured sphere and close one of the simple punctures.
In section \ref{M>2minimal} we study the $(M+2)$-punctured sphere with $M$ minimal punctures. Finally, in section \ref{TNk} we discuss the strong coupling limit and obtain the strongly coupled non-Lagrangian trinions $T_{N}^{k}$.

\section{Review}
\label{review}

In this section we review the necessary background, including what is known for the theories in class  $\mathcal{S}_k$.
We begin with the M-theory construction where implementing the orbifold is very simple (geometric), then continue with orbifolding on the gauge theory side.

\subsection{M-theory realization}
\label{sec:Setup}

\begin{table}[h!]
\centering
\begin{tabular}{|c|c|c|c|c|c|c|c|c|c|c|c|}
\hline
&$x^0$ & $x^1$ &$x^2$ &$x^3$ &$x^4$ &$x^5$ &$x^6$ &$x^7$ &$x^8$ &$x^9$ &($x^{10}$) \\
\hline
$M$ NS5 branes &$-$&$-$&$-$&$-$&$-$&$-$&.&.&.&.&.\\
\hline
$N$ D4-branes &$-$&$-$&$-$&$-$&.&.&$-$&.&.&.&$-$\\
\hline
$A_{k-1}$ orbifold&$.$&$.$&$.$&$.$&$-$&$-$&$.$&$-$&$-$&.&.\\
\hline
\end{tabular}
\caption{\it Type IIA brane configuration  for the 4D  $\mathcal{N}=1$ theories of
class $\mathcal{S}_k$.}
\label{configIIA}
\end{table}

The easiest way to introduce the theories in  class  $\mathcal{S}_k$  is to begin with the type IIA string theory brane setup in table
\ref{configIIA}, which was originally considered in \cite{Lykken:1997gy,Lykken:1997ub}.
Without imposing the $A_{k-1}$ orbifold we describe the $\cN=2$ theories  in  class  $\mathcal{S}$  \cite{Gaiotto:2009we}.
 The $SU(2)_R$ R-symmetry  of the $\cN=2$ theories corresponds to the  rotation symmetry of
$x^{7},x^{8}$ and $x^{9}$ and gets broken by the orbifold to the $U(1)_R$ symmetry of $x^{7},x^{8}$ rotations.
Rotation on the $x^{4},x^{5}$ plane corresponds to the $U(1)_r$ symmetry of the $\cN=2$ theories, which is preserved in the presence the orbifold singularity.

Following \cite{Witten:1997sc}, we wish to derive the SW curves using the uplift to M-theory of  table
\ref{configIIA}, and we define the holomorphic coordinates
\be
  v \equiv x^4 + i x^5 ~, \qquad s \equiv x^6 + i x^{10}  \qquad \mbox{and} \qquad
  w \equiv x^7 + i x^{8} ~
\ee
in terms of which we will write the spectral curves. It is also useful to define the exponentiated
\be
 t\equiv e^{-\frac{s}{R_{10}}} ~,
\label{ts}
\ee
where $R_{10}$ is the M-theory circle.  See \cite{Bao:2011rc} for the conventions we follow.
In order to account for the orbifold action, we impose
the identification
\begin{equation}
\label{OrbifoldAction}
\left( v  \,  , \,  w \right)  ~     \sim     ~ \left(  e^{\frac{2\pi i}{k}}v \,  , \, e^{-\frac{2\pi i}{k}}w  \right) \, .
\end{equation}
The coordinate $x^9$ is not part of a complex coordinate, which is consistent with
\cite{Lykken:1997gy,Lykken:1997ub}.

\bigskip

\begin{figure}[h]
\centering
\includegraphics[width=0.4\textwidth]{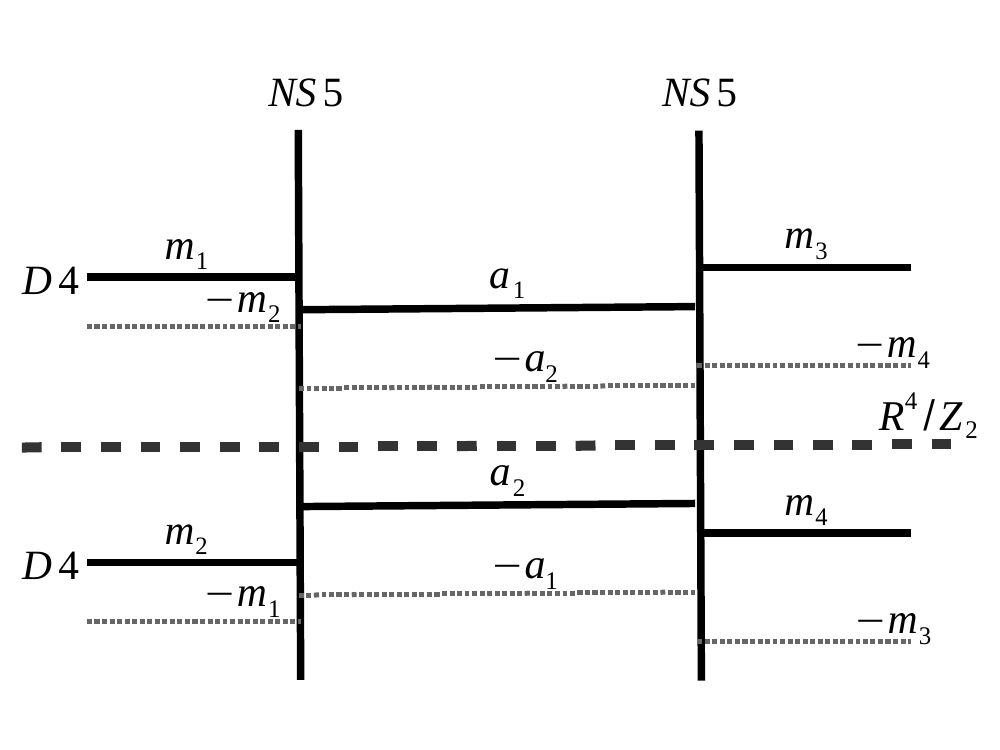}
\caption{\it The IIA brane set-up from which we calculate the IR curve $\Sigma$ for the $SU(2)$ case. The thick dashed line depicts the $\bZ_k$ orbifold point, for $k=2$. For each D4 brane the   mirror images are also depicted using grey dotted lines.}
\label{IIAbranesOrbifold}
\end{figure}

For $k=1$ using the set-up above we can obtain the spectral curve $\Sigma$, which is the spectral curve in the IR, by thinking of a single M5 brane with non-trivial topology and the appropriate boundary conditions (given by the asymptotic positions of the D4 and the NS5 branes) wrapping $\Sigma$.
Following Gaiotto  \cite{Gaiotto:2009we} for SCFTs, after a change of variables, we can rewrite the IR curve $\Sigma$ as a curve that describes $N$  M5 branes wrapping a different Riemann surface  $\mathcal{C}_{g,n}$ with genus $g$ and $n$ punctures, referred to as the Gaiotto curve or the UV curve.
We can equivalently think that $\mathcal{N}=2$ SCFTs are obtained  using M-theory  on $\mathbb{R}^{3,1}\times CY_{2} \times \mathbb{R}^{3}$ by wrapping M5 branes on the $CY_{2}$. The $ CY_{2} = T^* \mathcal{C}_{g,n}$ is the cotangent bundle over the UV curve. In this construction, we also have to specify the SW differential $\lambda_{SW}$ through
 \be
 \label{SWdifferential}
 \Omega_2 = dv\wedge ds = d\left( v ds\right) = d \lambda_{SW} \, .
 \ee
In M-theory  BPS states are described as  open M2 branes (the mass of which is computed by integrating
the holomorphic worldvolume two-form of the M2 brane) attached to an M5
brane whose volume is minimized.
The SW differential is obtained by solving the equation
 \be
 v^{N} = \sum_{\ell=2}^N  v^{N-\ell} \phi_{\ell}(t)
 \ee
 where $\phi_{\ell}(t)$ are meromorphic $\ell$-differentials with poles at the punctures and the meromorphic  sections
 of $\mathcal{L}_v^{\otimes \ell}$
have degree $\mbox{deg}\left(\mathcal{L}_v^{\otimes \ell}\right) = - \ell p$. This is related to  the genus of $\mathcal{C}_{g,n}$ as
$p=2(g-1)+n$.

 \bigskip

Following  \cite{Bah:2012dg,Xie:2013gma},   a large class of $\mathcal{N}=1$ theories can be obtained using M-theory on
$\mathbb{R}^{3,1}\times CY_{3} \times \mathbb{R}^{1}$, where the $CY_{3}$ is locally made out of two holomorphic line bundles on the curve
$\mathcal{L}_v \oplus \mathcal{L}_w   \longrightarrow   \mathcal{C}_{g,n}.
$\footnote{Sometimes also written as $\mathcal{O}(-p) \oplus \mathcal{O}(-q) \rightarrow   \mathcal{C}_{g,n}$.}  The holomorphic line bundles are such that their first Chern classes are $c_1\left(\mathcal{L}_v\right) = p$ and $c_1\left(\mathcal{L}_w\right) = q$.
The condition $c_1\left(CY_{3}\right) = 0$ leads to $p+q=2(g-1)+n$, which can also be
 reproduced from the field theory side \cite{Bah:2011vv,Bah:2012dg} using anomalies.
The  $\mathcal{N}=1$ spectral curve is an overdetermined algebraic system of equations \cite{Xie:2014yya,Giacomelli:2014rna}
\bea
\label{genericN=1curve}
&& v^{N} = \sum_{\ell=2}^N  v^{N-\ell} \phi_{v\ell}(t)
\\ \nonumber
&&  v^a w^b =  \sum_{i,j=2}^N c^{a,b}_{i,j}    \varphi_{i,j}(t,u)  v^{a-i} w^{b-j} \qquad \mbox{with} \qquad a+b=N\, , \quad a,b\geq 0
 \\ \nonumber
&& w^{N} = \sum_{\ell=2}^N  w^{N-\ell} \phi_{w\ell}(t)
\eea
with  $\phi_{a\ell}$ ($a=v,w$)  meromorphic  sections  of $\mathcal{L}_a^{\otimes \ell}$
 and $\varphi_{i,j}(t,u)$  meromorphic  sections  of $\mathcal{L}_v^{\otimes i}\otimes \mathcal{L}_w^{\otimes j}$. In the case of  $\mathcal{N}=1$ theories we have to specify the  holomorphic three-form
 \be
 \Omega_3 = dw\wedge dv\wedge ds   ~,
 \ee
whose integral gives us the holomorphic $\tau_{ij}W^{i\,\alpha} W^j_{\alpha}$ \cite{Witten:1997ep}.

 \bigskip

This construction simplifies greatly for $\mathcal{N} = 1$ SCFTs in 
\cite{Gaiotto:2015usa} class $\mathcal{S}_k$.
These are obtained as
 orbifolds of  $\mathcal{N}=2$ theories and thus M-theory on $\mathbb{R}^{3,1}\times \left( CY_{2}\times \mathbb{R}^{2}\right)/\mathbb{Z}_k \times \mathbb{R}^{1}$, with the orbifold action given in \eqref{OrbifoldAction}.
 In this case the $\mathcal{L}_w$ bundle is trivial and the
meromorphic  sections on $\mathcal{L}_w^{\otimes \ell}$ are global,  $c_1\left(\mathcal{L}_w\right) = 0$, so the second and the third lines in \eqref{genericN=1curve} will not play any role in our discussion  \cite{Lykken:1997gy,Lykken:1997ub}. Moreover, exactly because
$\mathcal{L}_w$ is trivial, $i.e.$ a direct product with   $T^* \mathcal{C}_{g,n}$,
the holomorphic three-form can be written as
\be
 \Omega_3 = f(w) dw\wedge  d \lambda_{SW}  ~,
 \ee
 with $f(w)\sim \sum_i \delta(w-w_i)$. The $w_i$ are the positions of the D4 branes on the $w$ plane, which we take to be all at the origin $w_i=0$.
This allows us integrate separately $\int  \Omega_3 =\int f(w) dw \int d \lambda_{SW} \sim \int d \lambda_{SW}$ and, up to an overall normalization that we drop, to just consider integrating  $\lambda_{SW}$ as for the theories with $\mathcal{N}=2$ supersymmetry.
According to \cite{Lykken:1997gy,Lykken:1997ub}, this remains true even when we resolve the orbifold.

\subsection{Field theory}

\begin{figure}[h]
\centering
\includegraphics[width=0.6\textwidth]{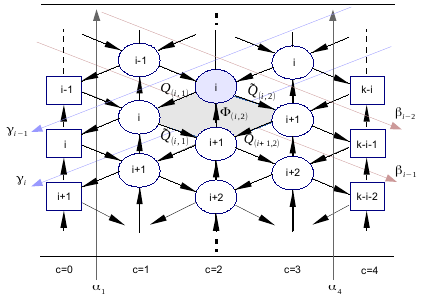}
\caption{\it The quiver diagram for the orbifolded linear quiver of $\cN=2$ with $M=4$.
The color groups are labelled by $(i,c)$ where ${i=1,\ldots,k}$ is the $\bZ_k$ orbifold index that labels the mirror images and $c=0,\ldots,M$ is the label from the original $\cN=2$ theory.}
\label{QuiverNodes1}
\end{figure}
The field theory side is understood following  \cite{Douglas:1996sw}.
We begin the study of the $\mathcal{N} = 1$ SCFTs in class $\cS_k$ by considering some theories with a Lagrangian description.
An important such family of theories is described by quiver diagrams depicted in
figure \ref{QuiverNodes1}. The ovals correspond to gauge groups and the
square boxes stand for flavor symmetries.  For $k=1$, these are the $\cN=2$ linear quivers
with superpotential
\be
\label{eq:N=2superpotential}
W_{\mathcal{S}} = \sum_{c=1}^{M-1}\left(
{Q}_{(c-1)}\Phi_{(c)} \tilde{Q}_{(c-1)}
-
\, \tilde{Q}_{(c)}\Phi_{(c)} Q_{(c)} \right) \, ,
\ee
where the index $c=1,\ldots,M-1$ labels the different color groups of the linear quiver,
obtained in IIA by
using $M$ NS5 branes. The nodes $c=0,M$ correspond to semi-infinite stacks of D4-branes.
The chiral field $Q_{(c)}$ corresponds to an arrow pointing left from node $(c+1)$ in
the quiver diagram to node $(c)$ and $\tilde Q_{(c)}$ corresponds to an arrow pointing
right from node $(c)$ to node $(c+1)$. The relative minus sign is crucial for preserving
$\mathcal{N}=2$ supersymmetry.

Imposing the $\mathbb{Z}_k$ orbifold breaks supersymmetry to $\cN=1$ and the superpotential becomes
\be
\label{eq:orbifoldedsuperpotential}
W_{\mathcal{S}_k} =\sum_{i=1}^k\sum_{c=1}^{M-1}
\left( {Q}_{(i,c-1)}\Phi_{(i,c)}\tilde{Q}_{(i,c-1)}  -
 \tilde{Q}_{(i,c)}\Phi_{(i,c)} Q_{(i+1,c)} \right) ~.
\ee
A chiral field $Q_{(i,c)}$ corresponds to an arrow pointing left into the node $(i,c)$
and $\tilde Q_{(i,c)}$ corresponds to an arrow pointing right from the node $(i,c)$.
The chiral field $\Phi_{(i,c)}$ points from $(i+1,c)$ to $(i,c)$.
The transformation properties of all the fields in the Lagrangian for the various gauge and global symmetries
are summarized in table \ref{tab:SymmetriesSCQCD}. In particular,
in class $\cS_k$ we have a large number of global $U(1)$ symmetries  \cite{Gaiotto:2015usa},
the action of which on the various bi-fundamental fields (arrows) is depicted by grey, blue and red arrows in figure \ref{QuiverNodes1}.

\begin{table}[h!]
 \centering{\small
    \begin{tabular}{| c || c | c | c | c | c | c | c |  }
    \hline
   & $SU(N)_{(i,c-1)} $  & $SU(N)_{(i,c)} $ & $SU(N)_{(i+1,c)} $   &   $U(1)_t$ &  $U(1)_{\alpha_c}$ &
$U(1)_{\beta_{i+1-c}}$   &   $U(1)_{\gamma_i}$ \\
    \hline
	 \hline
	$V_{(i,c)} $  &  $\bf{1}$    &  adj.  &   $\bf{1}$ & 0 & 0 & 0 & 0
	 \\
	 \hline
	 $\Phi_{(i,c)} $ &   $\bf{1}$ & $\Box$  &$\overline{\Box}$   &  $-1$  & 0 & $-1$ & $+1$
	 \\
	 \hline
	 $Q_{(i,c-1)} $  & $\Box$  & $\overline{\Box}$ & $\bf{1}$ &   $+1/2$ &  $-1$ & $+1$ & 0
	 \\
	 \hline
	 $\widetilde{Q}_{(i,c-1)} $ & $\overline{\Box}$ & $\bf{1}$ & $\Box$  & $+1/2$ & +1 & 0 & $-1$
	 \\
	  \hline
    \end{tabular} }
	 \caption{\it The symmetries of the fields of the orbifolded linear quivers in class $\mathcal{S}_k$. Apart from the color structure that can be read from the quiver diagram in figure \ref{QuiverNodes1}, there is a number of $U(1)$ global symmetries which can also be read from the extra arrows that intersect the bi-fundamentals.}
	 \label{tab:SymmetriesSCQCD}
\end{table}

\subsection{Coulomb moduli parameters and mass parameters}
From the study of $\mathcal{N}=2$ theories we are used to the fact that the SW curves are parameterized by the vacuum expectation values of gauge invariant operators
 $\langle  \mbox{tr} \,\phi^\ell \rangle \sim u_{\ell}$ that parametrize the Coulomb branch of the theory.
  There, solving the theory amounts to calculating the
vev $\langle \phi \rangle = a$ of the Coulomb moduli as a function of the $u$, $a(u)$, as well as their magnetic duals $a_D(u)$. This is done by computing the $A$- and $B$-cycle integrals, see  section \ref{ABintegrals} for a review,  and at weak coupling we find $a \sim \sqrt{u_2}$. The $a(u)$ in the IIA/M-theory picture correspond to the positions of the D4/M5 branes.

After orbifolding, the gauge invariant operators whose vevs parameterize the Coulomb branch of the theory are
 \footnote{More rigorously, the Coulomb moduli parameters $u$ which appear in the spectral curve in the later
sections are accompanied by
a certain linear combination of the product of these operators with the same total mass dimension
together with the correction from the mass parameters.
In \eqref{GaugeInvarOper} we omit these corrections and write the relation symbolically.}
\be
\label{GaugeInvarOper}
\langle  \mbox{tr} \left( \Phi_{(1,c)} \cdots \Phi_{(k,c)} \right)^\ell \rangle \sim u_{\ell k ,c} \, ,
\ee
where  the index $c=0,\ldots,M$ labels the different color groups in the IIA setup
with $M$ NS5 branes.
Notice that the bi-fundamental field $\Phi_{(i,c)}$ of the $\cN=1$ orbifold daughter theory is embedded into
the adjoint $\cN=2$ field $\Phi_{(c)}$ of the original (mother) theory \cite{Douglas:1996sw} as
\begin{eqnarray}
\Phi_{(c)}  = \left(
\begin{array}{ccccccc}
 & \Phi_{(1,c)}  & \\
 &  & \Phi_{(2,c)} & \\
& & & & \ddots  & & \\
 & & & & & & \Phi_{(k-1,c)}  \\
\Phi_{(k,c)} & & & \\
\end{array}
\right) ~.
\end{eqnarray}
If we diagonalize the vacuum expectation value of this field by a proper unitary matrix  $U$%
\footnote{The unitary matrix $U$ was included in the original $U(Nk)$ gauge transformation
but not in the $U(N)^k$ gauge transformation of the orbifolded theory.},
 we obtain
\begin{eqnarray}
\langle U^{-1} \Phi_{(c)} U  \rangle=
\mathrm{diag} \left( a_{(c)1}, a_{(c)2}, \cdots, a_{(c)N} \right)
\otimes \mathrm{diag}
\left( 1, e^{\frac{2 \pi i}{k}}, e^{\frac{4 \pi i }{k}}
\cdots
e^{\frac{2 \pi i (k-1)}{k}} \right) .
\label{Diag}
\end{eqnarray}
For $c=1,2, \cdots, M-1$,
the components $e^{\frac{2 \pi i n}{k}} a_{(c)I}$ of this matrix
are essentially what appear as Coulomb moduli parameters in the spectral curves in the rest of the sections.
For $c=0,M$, they are mass parameters, which will be denoted as $a_{(0)I} = m_I$ and $a_{(M)I} = m_{N+I}$
when we consider the case $M=2$ in later sections. These are the positions of the D4/M5 branes (all the mirror images) in the IIA/M-theory setup.

\section{The four-punctured sphere}
\label{4puncturedS}

In this section we construct the SW curve for the orbifolded $\mathcal{N}=2$ SCQCD  with $SU(N)$ gauge group and $N_f=2N$ flavors. We will refer to it as SCQCD$_k$. This is the class $\mathcal{S}_k$ generalization of the four-punctured sphere of Gaiotto.

\subsection{The curves}
\label{sec:curves}

Let us begin by recalling the SW curve of the $\mathcal{N}=2$ SCQCD with $SU(N)$ gauge group and $N_f=2N$ flavors.
Following Witten \cite{Witten:1997sc} this curve can be easily written by considering the M-theory uplift of the type IIA setup in table \ref{configIIA}.
This curve is derived based on the asymptotic behavior at large $v$ and is given by
\be
\label{genericSUNf}
\prod_{i=1}^N(v-m_i) ~ t^2~ + ~ (-(1+q)  v^N + q M  v^{N-1} + \sum_{\ell=2}^N
u_\ell v^{N-\ell})~t ~ + ~ q \prod_{i=N+1}^{2N}(v-m_i) ~ = 0 ~,
\ee
where $m_i$ with $i=1,\dots,N_f=2N$ denote the masses of the fundamental flavor hypermultiplets, $u_\ell$ with $\ell=2,\dots,N$ the Coulomb branch vacuum expectation values of $\langle \phi^\ell \rangle$ and $q=e^{2\pi i \tau}$ the UV coupling constant. The parameter $M=\sum_{i=1}^{2N}m_i$ is the sum of all the masses.
We moreover find it convenient to define the parameters $\fc^{(\ell)}_{m}$ for some parameter $m$ to be the singlet combinations or Casimirs of some symmetry, here the flavor symmetry,
\be
\label{CasimirDef}
\fc^{(\ell)}_{L} =\sum_{1=i_1<i_2<\ldots <i_\ell \leq N} m_{i_1} m_{i_2} \ldots
m_{i_\ell}
~,
\quad
\fc^{(\ell)}_{R} =\sum_{N+1=i_1<i_2<\ldots <i_\ell \leq 2N} m_{i_1} m_{i_2} \ldots
m_{i_\ell} ~,
\ee
in terms of which we write $M=\fc_L^{(1)} + \fc_R^{(1)}$.

To implement the orbifold we follow \cite{Lykken:1997gy,Lykken:1997ub}.
Orbifolding imposes  the identification
\be
v\,\sim\, e^{\frac{2\pi i}{k}} v ~.
\ee
For each mass $m_i$ there are $k$ mirror images on the $v$-plane and thus we must replace
\be
\label{orbiPrefRep}
(v-m_i)  \longrightarrow  \prod^{k}_{n=1} (v-m_i^{(n)})  =  (v^k-m_i^k)~.
\ee
The equality follows because also the (mirror images of the) mass parameters obey the orbifold condition and get identified as
\be
\boxed{
\label{orbifoldrestriction1}
m_i^{(n)}=  e^{\frac{2 \pi i n}{k}} m_i ~,\quad n=1,\ldots,k
}
\ee
Combining the replacement \eqref{orbiPrefRep} with equation \eqref{genericSUNf} gives
\bea
\label{N=2kCascurve}
\prod_{i=1}^{N}(v^k-m_i^k)t^2+P(v)t+q
\prod_{i=N+1}^{2N} (v^k-m_i^k)= 0    ~.
\eea
The polynomial $P(v)$ has degree $Nk$ because of the orbifold
\be
P(v) = -(1+q)v^{Nk} + u_{1}v^{Nk-1} + \cdots +  u_{Nk-1}v +  u_{Nk} ~,
\ee
but its monomials must respect the orbifold $\bZ_k$ symmetry,
as they must eventually be matched to the vevs of the gauge invariant operators
\eqref{GaugeInvarOper} that parameterize the Coulomb branch.\footnote{Note that in this paper we only study the Coulomb branch of the $\cS_k$ theories. We do not turn on vevs for the mesons or the baryons.}
 Any polynomial in $X=v^k$ will do that, so $P(v) = P_N(X)$ with
\be
P_N(X) = -(1+q)X^N+  \sum_{\ell=1}^{N} u_{\ell k}\,  X^{N-\ell}  ~.
\ee
Thus the spectral curve that describes the Coulomb branch of $SU(N)$ SCQCD$_k$ reads
\be
\label{SWcurveNS52Zk}
\boxed{
\prod_{i=1}^N (v^k - m_i{}^{k}) t^2+\left(-(1+q)v^{Nk}
+ \sum_{\ell=1}^N u_{\ell k} v^{(N-\ell) k} \right)t
+ q\prod_{i=N+1}^{2N} (v^k -m_i{}^{k}) = 0 ~.
}
\ee

\bigskip

\bigskip

\begin{figure}[t]
 \begin{center}
\includegraphics[width=10cm,]{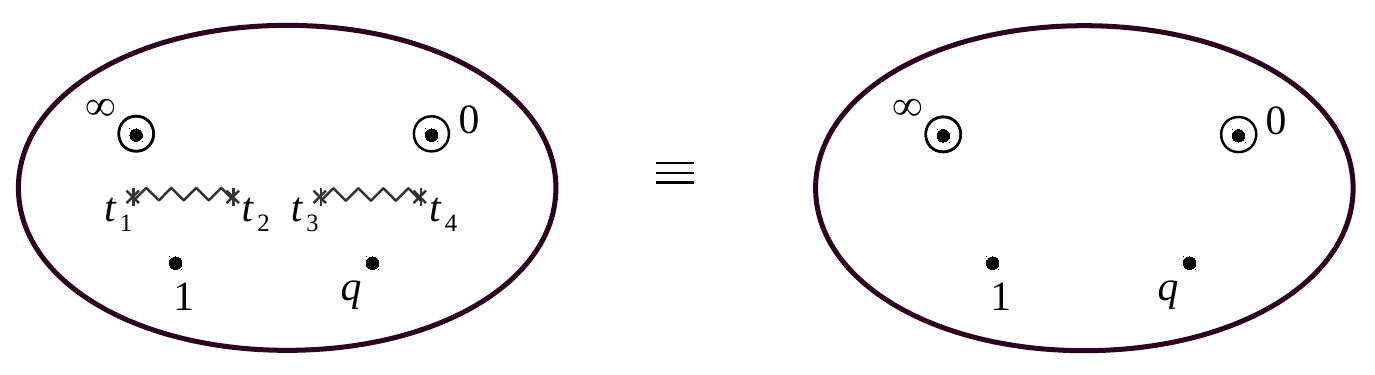}
 \end{center}
 \caption{ \it The spectral curve for $\mathcal{N}=2$ SCQCD with $SU(2)$ gauge group and $N_f=4$ flavors. On the left we depict the SW curve $\Sigma$.  On the right we depict the four-punctured sphere $\cC_{0,4}$,  the Gaiotto  curve, whose double-cover is $\Sigma$.}
 \label{fig:N=2case}
\end{figure}
We now want, following Gaiotto \cite{Gaiotto:2009we}, to rewrite this curve as the four-punctured sphere $\cC_{0,4}^{(k)}$ in class $\cS_k$. The first step in order to achieve this is to rewrite the
spectral curve \eqref{SWcurveNS52Zk}, which is  a polynomial in $t$, as a polynomial in $v$
\be
\label{Neq3si2NS5b}
v^{Nk}
+ \sum_{\ell=1}^N \frac{(-1)^{\ell}P_\ell(t)}{(t-1)(t-q)}v^{(N-\ell)k} = 0 ~,
\ee
where
\bea
P_\ell(t) = \fc_L^{(\ell,k)}t^2 + (-1)^\ell  u_k t+ q \fc_R^{(\ell,k)}
\eea
with  the  parameters $\fc_{L},~\fc_{R}$ defining the singlet combinations of the masses like in \eqref{CasimirDef}
\bea
&&\fc_L^{(\ell,k)} = \sum_{1 \le i_1 < i_2 < \cdots < i_\ell \le N}
m_{i_1}{}^k m_{i_2}{}^k \cdots m_{i_\ell}{}^k ~,
\\
&&\fc_R^{(\ell,k)} = \sum_{N+1 \le j_1 < j_2 < \cdots < j_\ell \le 2N}
m_{j_1}{}^k m_{j_2}{}^k \cdots m_{j_\ell}{}^k ~.
\eea
Then we make the substitution $v=xt$ in  \eqref{Neq3si2NS5b}, reparametrize
$t\rightarrow\frac{az+b}{cz+d}$  and
$x\rightarrow(cz+d)^2x$ and find\footnote{with  $\alpha = a-c$, $\beta = b-d$, $\zeta = a - cq$, $\xi = b - dq$.}
\be \label{Neq3si2NS5b2}
x^{Nk}=\sum_{\ell=1}^N\frac{\fp_{(\ell,2)}(z)}
{(cz+d)^{\ell k}(\alpha z+\beta)(\zeta z + \xi)(az+b)^{\ell k}} x^{(N-\ell)k}
\equiv
\sum_{\ell=1}^N
\phi_{\ell k}(z) x^{(N-\ell)k}
 ~.
\ee
This change of variables leaves invariant the SW differential for $ad-bc=1$.
Moreover, the $\fp_{(\ell,2)}(z)$ are degree two polynomials in $z$
\be
\fp_{(\ell,2)}(z) = (-1)^{\ell+1} (az+b)^2 \fc_L^{(\ell,k)} - (az+b)(cz+d) u_{\ell k} +
(-1)^{\ell+1}(cz+d)^2  q \fc_R^{(\ell,k)} ~.
\ee
Thus  $\phi_{\ell k}(z)$ are meromorphic  sections  of the line bundle $\mathcal{L}_v^{\otimes \ell k}$ of degree $-2\ell k$, $\mbox{deg}(\mathcal{L}_v)=-2$ and the space parametrized by $z$ is a four-punctured sphere\footnote{Recall that $p+q=2(g-1)+n$ and that for us $c_1\left(\mathcal{L}_w\right) =q= 0$.} $\cC_{0,4}^{(k)}$ in class $\cS_k$.

\bigskip

\subsection{Pole structure: Maximal and minimal punctures}
\label{sec:punctures}
Let us for simplicity begin with the $SU(2)$ SCQCD$_k$  theory.
As we will discover for the theories in class $\mathcal{S}_k$, the distinction between maximal and minimal punctures already appears for $N=2$.
This is in stark contrast with the $SU(2)$ punctures of  class $\mathcal{S}$ theories, which are indistinguishable\footnote{The  $SU(2)$ punctures in  class $\mathcal{S}$ are indistinguishable only after we shift $v$ appropriately, \eqref{GaiottoShift}.}.
The generalization of what we will do below to any $N$ is trivial for the maximal punctures.
The minimal punctures require more work and will be addressed in section \ref{SU(3)}.

We review shortly the $\cN=2$ case as a warmup.
The SW curve is obtained from \eqref{genericSUNf}
\be
\label{noOOrbiN2Nf4}
(v-m_1) (v-m_2)t^2 + \left(-(1+q)v^2+qMv +U\right)t+q (v-m_3)(v-m_4)=0 ~.
\ee
With this curve at hand, we look for its simple poles (positions of the punctures) and study its behavior close to them.  To do so we view the curve as a polynomial in $v$
\be
\label{keq1simplerev} (t-1)(t-q) v^2 -  P_1(t) v + P_2(t)   =0 ~ \\
\ee
with
\begin{subequations}
\begin{align}
\label{tpolynomialsP}
 P_1(t)& = (m_1+m_2)t^2 -q \, M\,t +q (m_3+m_4)  ~, \\
\label{tpolynomialsP2}
 P_2(t) & = m_1m_2 t^2  +u\,t  + q m_3m_4  ~.
\end{align}
\end{subequations}
and $M=m_1+m_2+m_3+m_4$. Solving for $v$ gives two solutions
\bea
\label{k1n2solX}
v_\pm = \frac{P_1(t) \pm \left(P_1(t)^2-4(t-1)(t-q)P_2(t)\right)^{1/2}}{2(t-1)(t-q)} ~,
\eea
which define a two-sheeted cover of a sphere parametrized by $t$.
At $t=1,q$ these become
\be
\label{vpmUnshiftefd}
v_{\pm_{~t= 1}} \sim
\left\lbrace \frac{m_1+m_2}{t-1}  ~,~ \frac{P_2(1)}{P_1(1)}\right\rbrace \quad\text{and}\quad
v_{\pm_{~t= q}} \sim
\left\lbrace -\frac{q(m_3+m_4)}{t-q}  ~,~ \frac{P_2(q)}{P_1(q)} \right\rbrace ~.
\ee
Consequently, $v$ has a simple pole on only one sheet close to $t=1,q$ and it is regular on the
other sheet. The residues are
\be
\text{Res}~ v_{\pm_{~t=1}} =
\left\lbrace m_1+m_2  ~,~ 0 \right\rbrace \quad\text{and}\quad
\text{Res}~ v_{\pm_{~t= q}} =
\left\lbrace -q(m_3+m_4)  ~,~ 0 \right\rbrace ~.
\ee
In the limits $t\rightarrow 0,\infty$ the solutions $v_\pm$ are
\be
v_{\pm_{~t\rightarrow \infty}} =\left\lbrace m_1 ~,~ m_2 \right\rbrace ~,\qquad
v_{\pm_{~t\rightarrow 0}} = \left\lbrace m_3 ~,~ m_4 \right\rbrace ~.
\ee

\paragraph{Gaiotto's shift:} It is possible to shift $v$ by a $t$-dependent function,
\be
\label{GaiottoShift}
\tilde v = v -\frac{1}{2}  \frac{P_1}{(t-1)(t-q)} ~,
\ee
such that  $\tilde v$ is the solution to
\be
\tilde v^2 = \frac{P_1^2 - 4(t-1)(t-q)P_2}{4(t-1)^2(t-q)^2}   \quad .
\ee
The SW differential $\lambda_{SW}$, as reviewed in section \ref{sec:Setup}, is given by the uniquely defined holomorphic two-from
\be
\label{SWdifferential}
\omega = ds \wedge dv = d \log t \wedge dv = d\left( v d \log t\right) = d \lambda_{SW}
\Longleftrightarrow \lambda_{SW}= v \frac{dt}{t}  + \mbox{const}(v)
\ee
where $\mbox{const}(v)$ means a constant with respect to $v$, which can depend on $t$.
The shifted $\lambda_{SW}$ in terms of $\tilde v$ has poles on both sheets.
Parametrizing $v=xt$, we finally find that the poles of the $SU(2)$ four-punctured
sphere are
\bea
x_{t= 0}&\sim&\frac{m_3-m_4}{2t} \left\lbrace +1,-1 \right\rbrace ~,~
x_{t=\infty}\sim\frac{t(m_1-m_2)}{2} \left\lbrace +1,-1 \right\rbrace~ \\
x_{t= 1}&\sim&\frac{m_1+m_2}{2(t-1)} \left\lbrace +1,-1 \right\rbrace~,~
x_{t= q}\sim-\frac{(m_3+m_4)}{2(t-q)} \left\lbrace +1,-1 \right\rbrace~.
\eea
The shift in $v$ leaves the physics unchanged\footnote{The two-form
$dv\wedge dt$ is invariant under the shift \eqref{GaiottoShift}.}
but reveals the full $SU(2)$ flavor symmetry of the punctures
at $t= 1,q$. The poles have residues which sum to zero. They have the
properties of an element of the Cartan subgroup of $SU(2)$ and thus get associated to its
fugacities, making the connection between the punctures and the $SU(2)$ flavor symmetries.

\bigskip

\paragraph{Back to class $\cS_k$:}  After performing the orbifold, the spectral curve becomes
\bea
\label{N=2kCascurve}
 (v^k-m_1^k)(v^k-m_2^k)t^2+P(v)t+q\prod^{k}_{n=1}
 (v^k-m_3^k)(v^k-m_4^k)= 0    ~.
\eea
When $k>1$, the curve \eqref{N=2kCascurve} has $2k$ solutions for $v(t)$, which are given by
\bea
\label{vksolutions}
v^{(n)}_{\pm}=e^{\frac{2\pi i n}{k}}v_{\pm} \quad \mbox{with} \quad
v^k_{\pm} = \frac{P_1(t)  \pm \sqrt{\Delta}}{2(t-1)(t-q)}
\eea
where $n=1\dots k$, $\Delta$ is the discriminant of the quadratic equation \eqref{N=2kCascurve} for $X =v^k$
\bea
\label{quadraticdiscriminant}
\Delta &=& (P_1(t))^2 - 4(t-1)(t-q)P_2(t)
\eea
and $P_{1,2}$ generalize the polynomials \eqref{tpolynomialsP}-\eqref{tpolynomialsP2}
\be
\label{P12}
P_1(t) = t^2\fc^{(1,k)}_L-u_kt+q\fc^{(1,k)}_R ~, \qquad P_2(t) =  t^2\fc^{(2,k)}_L+u_{2k}t+q\fc^{(2,k)}_R ~.
\ee
Let us begin by looking at  \eqref{vksolutions} close to $t=0$, where
\be
v^k_{\pm_{~t= 0}}=\left\lbrace m_3^k~,~m_4^k\right\rbrace ~~~\Rightarrow ~~~
\label{solvt0n2genk}
v^{(n)}_{\pm_{~t= 0}}=\left\lbrace m_3^{(n)},m_4^{(n)}\right\rbrace~
\ee
for $m_i^{(n)}$ introduced in \eqref{orbifoldrestriction1}.
Similarly, at $t\to \infty$, $v$ takes values
\be
\label{solvtinftyn2genk}
v^{(n)}_{\pm_{~t\to \infty}} = \left\lbrace m_1^{(n)},
m_2^{(n)}\right\rbrace   ~.
\ee
These are the maximal punctures of the curve parameterized by $t$ in class $\cS_k$.
At these punctures, the differential $\lambda_{SW}$ has a {\it simple pole  on all $2k$ sheets} of the spectral curve.
 The maximal punctures are  parameterized by $k$ mirror images of $U(2)$.
The generalization to the $SU(N)$ case is immediate
\begin{align}
& \lim_{~t\rightarrow \infty}   v^{(n)}_{1,\dots , N} =
\left\lbrace m_1^{(n)}, m_2^{(n)},\dots, m_N^{(n)}\right\rbrace   ~,
\\
& \lim_{~t\rightarrow 0}   v^{(n)}_{1,\dots , N} =
\left\lbrace m_{N+1}^{(n)}, m_{N+2}^{(n)},\dots, m_{2N}^{(n)}\right\rbrace   ~.
\end{align}
We thus have $k$ mirror images of  $U(N)$.

\bigskip

At this point it is important to discuss the following. For $\cN=2$ theories in class  $\cS$, we shifted $v$ to $\tilde{v}$ removing the $U(1)$ part (the sum of the masses) from the maximal puncture. For $\cN=1$ theories in class $\cS_k$  shifting $v$ is not possible any more! The orbifold breaks translational invariance on the $v$-plane and the origin $v=0$ is fixed on the orbifold point.

\bigskip

Close to $t=1,q$, the  behavior of the curve changes to
\bea \label{rest1n2noshift}
v_+^k~_{t\to 1} \sim \frac{1}{t-1} \frac{P_1(1)}{1-q} -\frac{P_2(1)}{P_1(1)} ~,
\qquad v_-^k~_{t\to 1} \sim \frac{P_2(1)}{P_1(1)} ~,\\
\label{restqn2noshift}
v_+^k~_{t\to q} \sim \frac{1}{t-q} \frac{P_1(q)}{q-1} -\frac{P_2(q)}{P_1(q)} ~,
\qquad v_-^k~_{t\to q} \sim \frac{P_2(q)}{P_1(q)} ~.
\eea
These are the minimal punctures of class $\cS_k$. Note first of all that they do not correspond to simple poles of the SW differential \eqref{SWdifferential}, but to branch points as
$v_+ \sim \left(t-t_\bullet \right)^{-1/k}$ with  $t_\bullet=1,q$.
Moreover, let us stress that it is not possible to make a closed curve around them staying on one sheet. Only when we dive through the cuts to all the different sheets we can form a closed loop, and the integral of $\lambda_{SW}$ along this loop gives zero.

The fact that $v_-$ is finite on $t_\bullet=1,q$ is a consequence of the fact that we cannot shift $v$, and is the same as for the $\cN=2$ case in equation \eqref{vpmUnshiftefd}.
The fact that the integral along the closed loop around the minimal puncture  is zero is a novel feature of class $\cS_k$, the physics of which is not
fully elucidated.
To understand the minimal punctures better we study the cut structure on the curve.


\subsection{Cut structure}
\label{sec:cuts}

\subsubsection{The $SU(2)$ case}

\begin{figure}[h!]
 \begin{center}
\includegraphics[width=10cm]{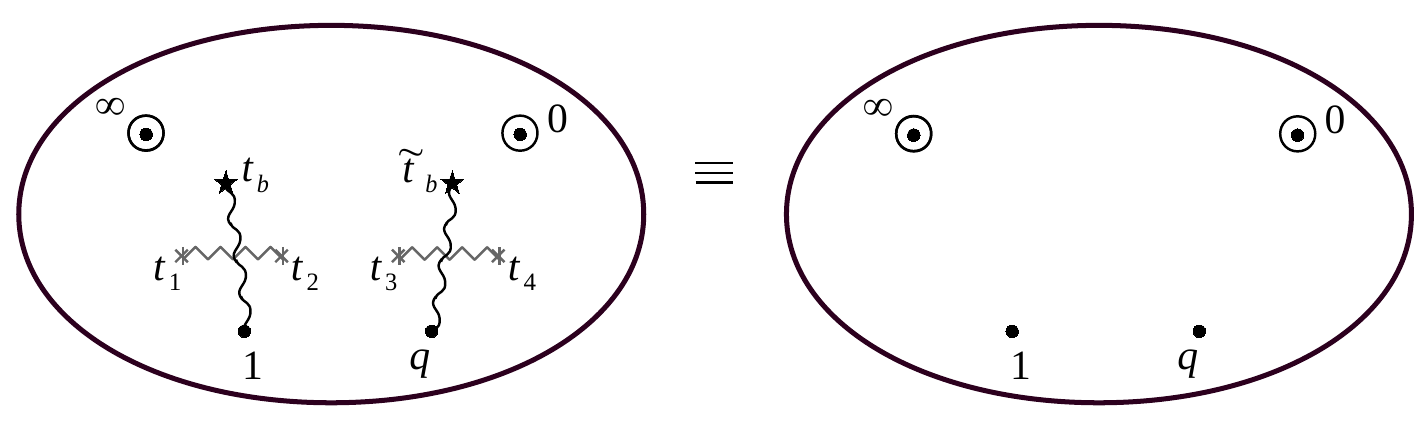}
 \end{center}
 \caption{ \it The IR spectral curve of $\mathcal{N}=1$ $SU(2)$ SCQCD$_k$ obtained as a $2k^{th}$ cover of the four-punctured sphere in class $\cS_k$. On the left we draw the IR curve $\Sigma$ and explicitly denote the branch cut structure. On the right side we depict only the four-punctured sphere $\mathcal{C}_{0,4}^{(k)}$.}
 \label{fig:Gaiotto}
\end{figure}

After having discussed the pole structure of the curve, we turn to the study of branch cuts.
As we will immediately demonstrate, our orbifolded curves have two types of branch
cuts. They have the cuts that are inherited from the parent $\cN=2$
theory and moreover they have novel branch cuts that appear due to the orbifold. The cut structure can be seen in figure \ref{fig:Gaiotto}.
We will first present everything for the $SU(2)$ SCQCD$_k$, the spectral curve of which  \eqref{N=2kCascurve} is a $2k$ polynomial in $v$.

\paragraph{Branch cuts inherited from the $\cN=2$ theory:}
The $2k$ solutions of \eqref{N=2kCascurve}  can be separated into the following two types \eqref{vksolutions}
\bea
\label{genk2solvk}
v^k_{\pm} = \frac{P_1(t)  \pm \sqrt{\Delta}}{2(t-1)(t-q)} ~, \qquad \Delta &=& (P_1(t))^2 - 4(t-1)(t-q)P_2(t) ~.
\eea
The $2k$ branches covering the $t$-sphere therefore comprise $k$ plus-type sheets and
$k$ minus-type sheets, which are joined by branch cuts.
The branch points are given by loci on the $t$-sphere where the solutions $v^k_{\pm}$ become
equal, which are the four solutions $t=t_i$, $i=1,\ldots, 4$ of $\Delta(t_i)=0$.
For later convenience, we fix the convention of $t_i$ in appendix \ref{app:B}.

\paragraph{Novel branch cuts due to the orbifold:}
By computing the asymptotic behavior of the
curve \eqref{N=2kCascurve}, we can see that the solutions in $v$ near $t=1$ and $t=q$ are \eqref{rest1n2noshift} and \eqref{restqn2noshift}
\begin{align} \label{specialpoints1}
&v_{+}^k \sim \frac{1}{t-t_\bullet}
 \textrm{ near }  t_\bullet=1,q,
\\
&v_{-}^k=\textrm{non-singular}   \, \textrm{ at } \, t_\bullet=1,q \, .
\end{align}
 Therefore the $v_{-}$ solution does not have a pole
associated with the simple puncture nor a branch cut.
On the one hand, the $v_{+}$ sheets develop singularities ${(t-t_\bullet)^{-1/k}}$ and have
branch cuts. To find similar branch cuts for the $v_{-}$ solution, let
$t=t_b, \tilde t_b$ be two  solutions of $P_2(t)=0$, where  $P_2(t)$ is defined in \eqref{P12}.
These are
\be \label{VminusBP}
t_b=\frac{-u_{2k}-\sqrt{u_{2k}^2-4q \fc_L^{(2,k)}\fc_R^{(2,k)}}}{2\fc_L^{(2,k)}} ~,
\qquad
\tilde t_b=\frac{-u_{2k}+\sqrt{u_{2k}^2-4q \fc_L^{(2,k)}\fc_R^{(2,k)}}}{2\fc_L^{(2,k)}} ~.
\ee
Near one of these points $t = t_b$, $v^k_{\pm}$ behave as
\begin{align} \label{specialpoints2}
v_{+}^k \sim \frac{P_1(t_b)}{(t_b-1)(t_b-q)}, \qquad
v_{-}^k
 \sim \frac{\fc_L^{(2,k)} (t_b-\tilde{t}_b)(t-t_b)}{P_1(t_b)}
\end{align}
and similarly around $t=\tilde t_b$ by exchanging $t_b$ and $\tilde t_b$.
The two solutions of $P_2(t)=0$
create two branching points on the $v_{-}$ sheets as $v_{-}\sim (t-t_b)^{1/k}$.
The complete system of cuts is illustrated in figure \ref{fig:Gaiotto}.

We summarize all the possible distinguished points in table \ref{legend-points} and depict them on the spectral curve in  figure \ref{fig:Gaiotto}, along with the cut structure.
\begin{table}[h!]
\centering{\renewcommand{\arraystretch}{1.5}
\begin{tabular}{|c|c|c|}
\hline
$\odot$ & Maximal puncture &  $(t-t_\odot)^{-1}$  \\
\hline
$\bullet$ &  Minimal puncture &  $(t-t_\bullet)^{-\frac{1}{k}}$\\
\hline
$\bigstar$ & Novel branching point $t_b$ &  $(t-t_b)^{\frac{1}{k}}$\\
\hline
$\ast$ & $\mathcal{N}=2$ branching point $t_i$ &  $(t-t_i)^{\frac{1}{2}}$\\
\hline
\end{tabular}   }
\caption{\it All the distinguished points on the spectral curve of SU(2) SCQCD$_k$.}
\label{legend-points}
\end{table}

Schematically, what we have discovered is the following
\be
\lambda_{SW} =  \left(   \sum_\odot \frac{m_\odot}{z - z_\odot} +   \sum_\bullet \frac{\mu_\bullet}{ \sqrt[k]{z - z_\bullet}} + \dots  \right) dz
~.
\ee
This equation should be understood as follows: $\lambda_{SW}$ is not globally defined on the sphere $\mathcal{C}_{0,4}^{(k)}$. The domain of $z$ is taken around each singularity from the correct sheet (patch). The differential $\lambda_{SW}$ is  globally defined only on the spectral curve in the cotangent bundle over the UV curve $T^* \mathcal{C}_{g,n}$; only on the IR curve $\Sigma$.

\subsubsection{The $SU(3)$ case}
\label{SU(3)}

\begin{figure}[h!]
 \begin{center}
\includegraphics[width=12cm]{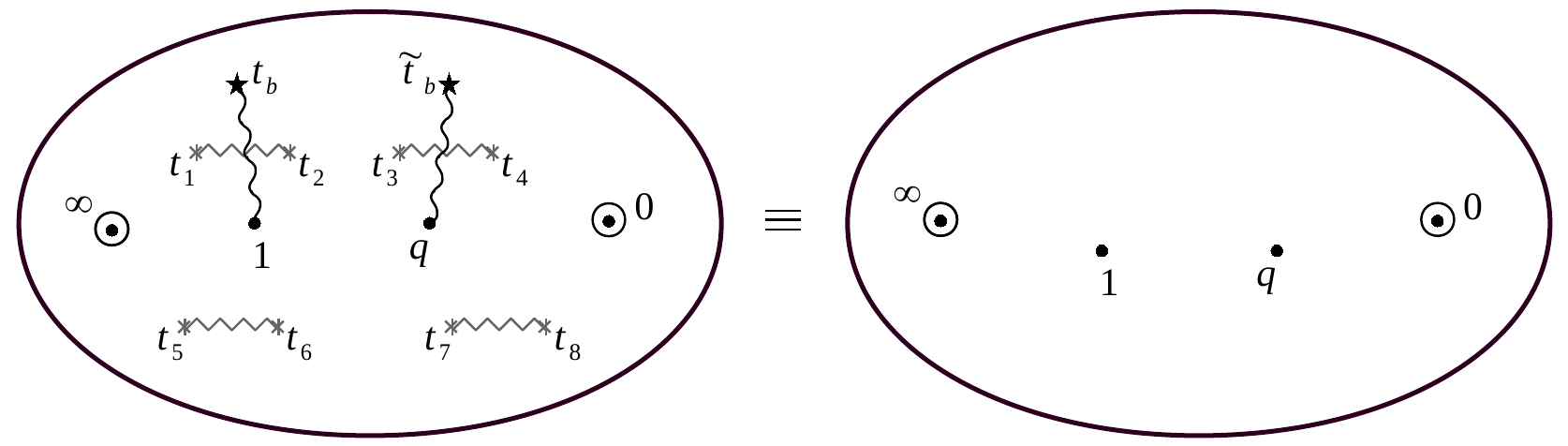}
 \end{center}
 \caption{ \it The IR spectral curve of $\mathcal{N}=1$ $SU(3)$ SCQCD$_k$ obtained as a $3k^{th}$ cover of the four-punctured sphere in class $\cS_k$. On the left we draw the IR curve $\Sigma$ and the branch cuts and on the right the four-punctured sphere $\mathcal{C}_{0,4}^{(k)}$.}
 \label{fig:SU3Nf6curve}
\end{figure}

Let us now turn to  the case of $SU(3)$ SCQCD$_k$.
The spectral curve takes the form
\begin{align}
\label{SU3eq}
(t-1)(t-q)v^{3k} - P_1(t)v^{2k}+P_2(t)v^k - P_3(t)=0
\end{align}
The $3k$ solutions 
can be separated into the following three types $v_{(i=1,2,3)}$
\be \label{SU3sol1}
3(t-1)(t-q)\left(v_{i}\right)^k
=P_1(t)
-\frac{\omega_i}{2^{1/3}}{\left(\sqrt{ \tilde\Delta(t) }-f_6(t)\right)^{1/3}}
+\frac{2^{1/3}f_4(t)}{\omega_i~\left(\sqrt{ \tilde\Delta(t) }-f_6(t)\right)^{1/3}} ~,
\ee
where $\omega_i=\omega^{2i}$ with $\omega=\frac{1+i\sqrt{3}}{2}$,
\be
\label{CubicDiscriminant}
\tilde\Delta(t)  = 4(f_4(t))^3+(f_6(t))^2
\ee
is  the discriminant of the cubic equation for $v^k$ up to an overall $(t-1)^2(t-q)^2$ factor and
$f_n(t)$ are order $n$ polynomials
\begin{align}
&f_4(t)=-(P_1(t))^2+3(t-1)(t-q)P_2(t),\\
&f_6(t)= 2(P_1(t))^3 - 9(t-1)(t-q)P_1(t)P_2(t) + 27(t-1)^2(t-q)^2P_3(t) \, .
\end{align}
These solutions describe the covering structure of 
four-punctured sphere $\cC_{0,4}^{(k)}$ with coordinate $t$. 
Below we study the branching structure of the resulting spectral curve, which is depicted in figure \ref{fig:SU3Nf6curve}.

\paragraph{Branch cuts inherited from the $\cN=2$ theory:}
 The branch points are given by loci on the $t$-sphere where two of the three  solutions $v^k_{(1,2,3)}$ become  equal.
 These points are precisely the zeros of the discriminant \eqref{CubicDiscriminant} of the cubic equation (\ref{SU3eq}) in $v^k$, $\tilde\Delta(t_j)=0$.
They create  branch cuts which connect pairwise the sheets  $v^k_{(i)}$  defined by the three solutions  \eqref{SU3sol1}. 
The discriminant of the cubic equation  has eight  zeros
\begin{align}
4(f_4(t))^3+(f_6(t))^2 =  (t-1)^2(t-q)^2  \prod_{j=1}^8(t-t_j)
\end{align}
 which we denote as $t=t_j$, $j=1,\ldots, 8$.
These eight  branching points are therefore intersection points of the three types of sheets.

\paragraph{Novel branch cuts due to the orbifold:} Let us now study the asymptotic behavior of the solutions in $v$ in the vicinity of $t=1,q$.
At these points, the solutions become very simple because polynomials reduce to
$f_4(t)=-(P_1(t))^2$ and
$f_6(t)=2(P_1(t))^3$ and
\begin{align}
&v_{2,3}(t)=\textrm{non-singular}\textrm{ at } t_\bullet=1,q,
\\
&
v_{1}(t)\sim\frac{P^{1/k}_{1}(t_\bullet)}{(t-t_\bullet)^{1/k}}\textrm{ near } t_\bullet=1,q,
\end{align}
since $1-\omega-1/\omega=0$.
The $v_{1}$ sheets therefore develop singularities and branch cuts ${(t-t_\bullet)^{-1/k}}$.
The other $2k$ sheets of different type do not have branching at these points.

We also have the special points $t=t_b,\,\tilde{t}_b$ on the curve that are solutions to the equation $P_3(t)=0$.
At these points, for example $t=t_b$, the original equation for the curve  reduces to
\be\label{su3zeros}
v^k\left((t_b-1)(t_b-q)v^{2k} - P_1(t_b)v^{k}+P_2(t_b)\right)=0
\ee
and similarly for $t=\tilde t_b$. Thus $k$ of the solutions become
\begin{align}
\left(v_{2}\right)^k=0 ~.   
\end{align}
On the sheets $v_{2}$ the curve then develops branching points of the type
\be
v_{2}\sim (t-t_b)^{1/k} ~.
\ee
Note that the $v_{3}$ sheets do not have any further branching points introduced by the orbifold.
This is because only $k$ solutions are zero in equation \eqref{su3zeros} (unless $P_2(t_b)=0$). 

\subsection{The Genus of the IR curve}
\label{sec:genus}

To extract physical information from these curves we need to compute the $A$- and $B$- cycle integrals. This is done in the next section.
Before however, we need to know the independent cycles of the spectral curve \eqref{SWcurveNS52Zk}.
The genus is computed via the Riemann-Hurwitz formula
\be
g(\Sigma) = 1 - Nk +\frac{b}{2} ~,
\ee
where $Nk$ is the  degree of  \eqref{SWcurveNS52Zk} as a polynomial in $v$, $b$ is the branching index
\be
b=\sum_{v_n} \left( \nu(v_n)-1 \right) ~,
\ee
$v_n$ are the ramification points on $\Sigma$ and $\nu(v_n)$ is the ramification index at $v_n$, which is the number of sheets that meet at $v_n$.
First for the $N=2$ curve in \eqref{N=2kCascurve} we have
four branching points $t_i$. They
correspond to $4k$ ramification points $v_s(t_i)$ with $\nu(v_s(t_i))=2$ and branching index $b=4k$.
The four additional branching points that appear because of the orbifold, $t_\bullet =1,q$ and $t_\bigstar=t_b,\tilde{t}_b$,
correspond to $v$ with ramification index $k$ and $b=2(k-1)$. The genus of the curve \eqref{N=2kCascurve}  is thus $2k-1$.
Then, the $N=3$ curve \eqref{SU3eq} is a  degree $3k$ polynomial in $v$ and has
$8k$ points with ramification index $2$ and four points with ramification index $k$. Thus, its genus is $3k-1$.
Finally, for arbitrary $N$ the genus of the curve \eqref{SWcurveNS52Zk}  is $Nk-1$.

\subsection{Various cycles and their integrals}
\label{ABintegrals}

In this section we wish to clarify the structure of the non-trivial cycles in the spectral curve and their relation to the physical parameters.

\begin{figure}[b]
\centering
\includegraphics[width=0.45\textwidth]{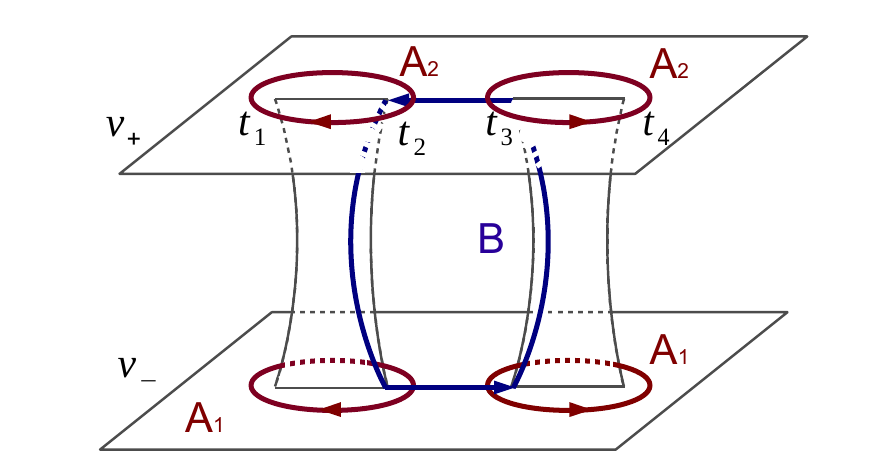}
\caption{\it For the $\cN=2$  theory with  $SU(2)$, $N_f=4$ and with $m_i=0$ $\forall \, i=1,\dots,4$ we depict here the  2-sheeted cover of the four-punctured sphere with the $A$- and $B$-cycle defined by the red and the blue contours.}
\label{2sheetedC2}
\end{figure}

Let us begin with a rapid review of  the SW solution for the $\cN=2$  theory with  $SU(2)$ and $N_f=4$: the moment the SW curve \eqref{noOOrbiN2Nf4} and the SW differential \eqref{SWdifferential} are known, we can immediately obtain
\be
a(u)=\frac{1}{2\pi i}\int_{A}\lambda_{SW}  ~~, \qquad a_D(u)=\frac{1}{2\pi i}
\int_{B}\lambda_{SW}
\qquad\mbox{and}\qquad
\tau =\frac{\partial a_D}{ \partial a}
\ee
where the $A$- and $B$- cycles are depicted in figure \ref{2sheetedC2}.
Note that $A$-cycle can be equivalently defined either as the cycle around the cut  $(t_1,t_2)$ or   $(t_3,t_4)$ up to  cycles around the poles at $t=0,q,1,\infty$. In the massless limit, where
the poles do not give any masses as their residues,
these two definitions become identical.
For future comparison, we also define $A_1$- and  $A_2$- cycles on the upper and lower sheets, which are equivalent here ($\cN=2$ case).
The relation $a_1(u)=-a_2(u)$ reflects the difference of the sign of $\lambda_{SW}$ on the upper and lower sheets.
This condition $a_1=-a_2 = a$ defines the zero on the $v$-plane and can always be obtained since we have the freedom to shift $v$, as discussed in section \ref{sec:punctures}. Using \eqref{k1n2solX} and \eqref{SWdifferential}, in the weak coupling and massless limit, we find
\be
a =  \sqrt{u}   \qquad \mbox{and} \qquad a_D  = \frac{1}{\pi i} \sqrt{u} ~
\mbox{log}(q)
\ee
which leads to
\be
\tau = \frac{\partial a_D}{\partial a}=\frac{1}{\pi i}~\mbox{log}(q)~.
\ee

\bigskip

\begin{figure}[b]
\centering
\includegraphics[width=0.6\textwidth]{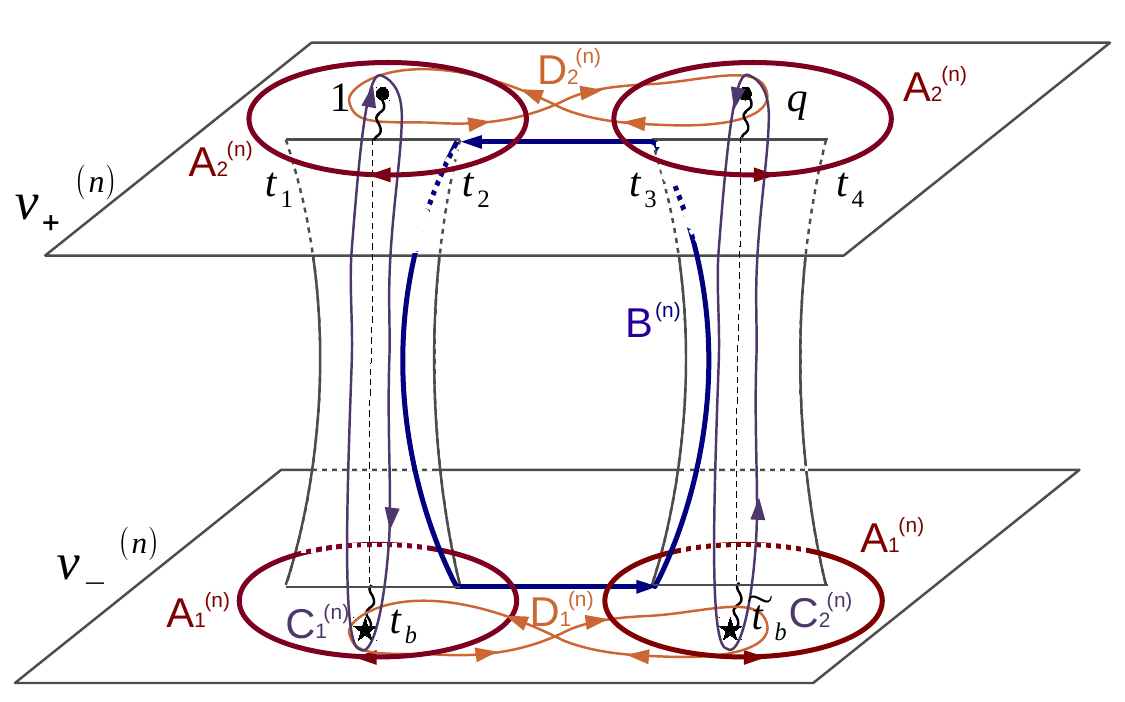}
\caption{\it The various cycles, including the novel $C$ and $D$ cycles, on the spectral curve of the $SU(2)$ SCQCD$_k$ theory in class $S_k$.}
\label{AnBncycle}
\end{figure}

\begin{figure}
\centering
\includegraphics[width=1.0\textwidth]{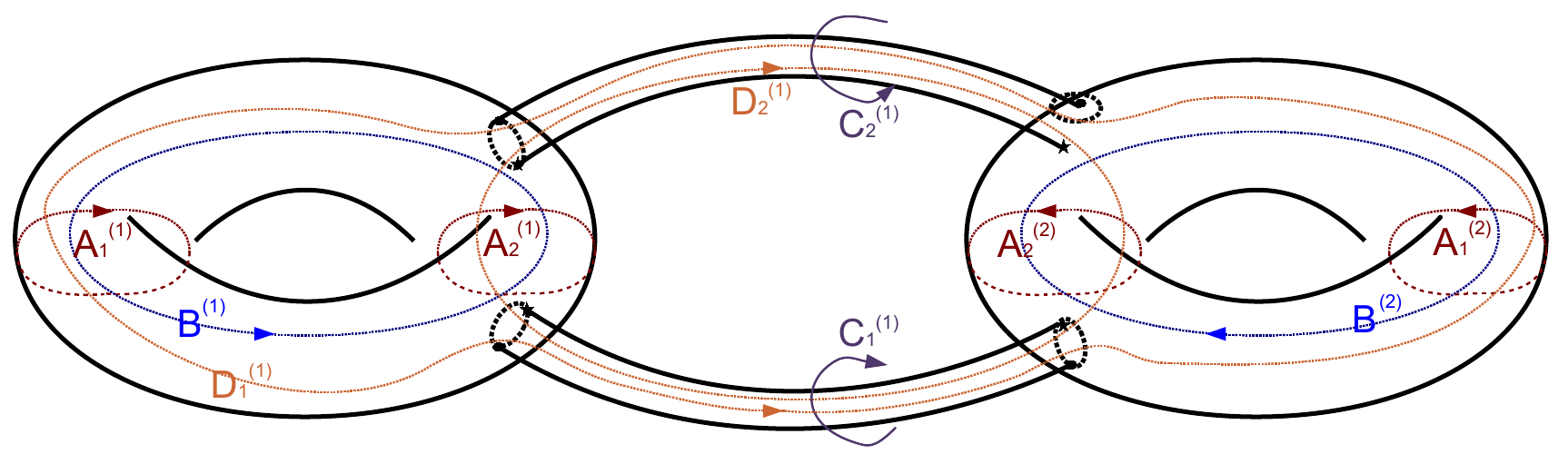}
\caption{\it The IR $SU(2)$, $k=2$ curve and its cycles: those in red and blue generalize the A- and B-
cycles of the $\cN=2$ $SU(2)$ $N_f=4$ theory. Those in purple and orange are due to the $\bZ_2$ orbifold.}
\label{Z2Torus}
\end{figure}

Let us now turn to the class $\cS_k$ theories. For simplicity we will confine ourselves to the $SU(2)$ case.
We can define various non-trivial cycles as depicted in figure \ref{AnBncycle}.
The $A_i^{(n)}$ $(i=1,2)$ and $B^{(n)}$ cycles are the counterparts of
$A$-cycles and $B$-cycles in the above mentioned $\cN=2$ theory.
On top of that, due to the novel branch cut described in section \ref{sec:cuts},
there exist cycles which we denote as $C^{(n)}_i$ and $D^{(n)}_i$ $(i=1,2)$.
All these new cycles have the label $n=1,\cdots, k$,
which indicates the $n$-th sheets on which $v_{\pm}^{(n)}$ are defined.
Although we have defined $7k$ cycles in total,
these are not all topologically independent
but satisfy the following relations for $n=1,\cdots ,k$
\bea
&&
C_1^{(n)} = C_2^{(n)}
= A_1^{(n)} + A_2^{(n)},
\nn \\
&&
B^{(n)} - B^{(n+1)} =
C_1^{(n)} + C_2^{(n+1)}+ D_1^{(n)} + D_2^{(n)},
\nn \\
&& \sum_{n=1}^k C_i^{(n)} = \sum_{n=1}^k D_i^{(n)} = 0
\label{top_rel}
\eea
up to the cycles around the singularities at $t=0,q,1,\infty$,
where $n=k+1$ should be understood as $n=1$.
The relations above are still redundant, as we actually have $3k+2$ independent constraints.
Therefore, we obtain $4k-2$ independent cycles. They generate all the non-trivial cycles in
the spectral curve, which has genus $2k-1$.

In order to understand  figure \ref{AnBncycle} more intuitively and the relations (\ref{top_rel})  graphically,
we concentrate on the case $k=2$ in figure \ref{Z2Torus}.
The structure of the spectral curve can be understood as follows:
first, due to the branch cuts inherited from the $\cN=2$ theory, we obtain a torus.
Due to the novel branch cuts,
we have $k=2$ copies of the $\cN=2$ tori connected to each other at the branch cuts
and we obtain the genus 3 Riemann surface depicted in figure \ref{Z2Torus}.
All the cycles defined above are shown in this figure and
we see that all of the relations in (\ref{top_rel}) are satisfied.
Their generalization to $k>2$ is also straightforward.

Analogous to the $\cN=2$ theory, the integrals around the
$A_i^{(n)}$ and $B^{(n)}$ cycles give
\be
a_i^{(n)}=\frac{1}{2\pi i}\int_{A_i^{(n)}}\lambda_{SW}\qquad\mbox{and}\qquad a^{(n)}_D=\frac{1}{2\pi i}
\int_{B^{(n)}}\lambda_{SW} ~.
\ee
Note that in general $a_1^{(n)} + a_2^{(n)}$ no longer vanishes.
From the relation \eqref{top_rel}, this is equal to the integral around
$C^{(n)}$ cycles
\be
\label{CcycleIntegral}
\frac{1}{2 \pi i} \int_{C^{(n)}} \lambda_{SW} = a_1^{(n)} + a_2^{(n)}~.
\ee
The information of the $D_i^{(n)}$ cycle integrals is available from \eqref{top_rel} and is given by
\be
\label{DcyclesIntegrals}
\frac{1}{2 \pi i} \int_{D^{(n)}_1} \lambda_{SW}
+ \frac{1}{2 \pi i} \int_{D^{(n)}_2} \lambda_{SW}
= ( a_D^{(n)} + a_D^{(n+1)} ) - (a_1^{(n)} + a_2^{(n)}) - (a_1^{(n+1)} + a_2^{(n+1)}).
\ee

Let us now  compute the integrals around the $A$- and $B$- cycles
in the weak coupling and massless limit.
For the $A^{(n)}_i$ cycles, the integrals around the cut $(t_3,t_4)$ are
approximated by the contour integrals around $t \sim q \sim 0$.
From equation \eqref{genk2solvk} and the SW differential \eqref{SWdifferential}, we obtain
\be
 \label{aparametersv2}
 a_1^{(m)}  = e^{\frac{2\pi i m}{k}}  \frac{ \left[-u_k - \sqrt{u_{k}^2+4u_{2k}}\right]^{1/k}}{2^{1/k}}
 \quad \mbox{and} \quad
  a_2^{(n)}  = e^{\frac{2\pi i n}{k}} \frac{ \left[-u_k + \sqrt{u_{k}^2+4u_{2k}} \right]^{1/k}}{2^{1/k}}
\ee
We then note that
\be
a_{1,2}{}^k=\frac{ - u_k \mp \sqrt{u_{k}^2+4u_{2k}}}{2} ~,
\ee
which leads to the weak coupling relation between the Coulomb moduli parameters
\be
\boxed{
\label{uka1a2}
  u_k = - (a_1^k+a_2^k)
   \qquad \mbox{and} \qquad
    u_{2k} =-a_1^k a_2^k ~.~
    }
\ee
Unlike the original $\mathcal{N}=2$ theory, now the combination $a_1^k+a_2^k$ is not vanishing.
Since the orbifold fixed point exists at $v=0$,
there is no freedom to shift $v$ without also changing this.
In general, the center of mass of the two D4-branes can be different from the orbifold fixed point
and thus we have one more Coulomb moduli parameter $u_k$.

The $B^{(n)}$ cycle integrals diverge as $\log q$ in the weak coupling limit,
due to the contribution of the integral close to the point $t = t_3 \sim q \sim 0$.
Here, we estimate only the diverging coefficient and ignore the finite part.
For this purpose, it is enough to approximate the integration range $(t_2,t_3)$ by $(1,q)$.
Also, we estimate the value of $v_{\pm}^{(n)}$ at $t \sim 0$,
which is given by $a_i^{(n)}$ as computed in equation  \eqref{aparametersv2}.
Therefore, we obtain
\bea
a_D^{(n)}
&\sim&\frac{e^{\frac{2\pi i n}{k}}}{2\pi i}\text{log}(q)\left(a_1-a_2
\right)~,\qquad n=1,\ldots,k~.
\eea
From the way in which the $2k$ sheets are pairwise connected, the choice of each
contour to run on the two sheets in a pair connected by a branch cut gives
\be
\tau^{(n,n)}=\frac{\partial a_D^{(n)}}{\partial a_1^{(n)}  } = - \frac{\partial a_D^{(n)}}{\partial a_2^{(n)}} =  \frac{1}{2\pi i}\text{log}(q) = \tau
\ee
all equal for $n=1,\ldots , k$.
In the weak coupling limit all the YM coupling constants are equal, as we are considering the theory on the orbifold point. In order to go away from the orbifold point, the orbifold must be replaced  by a $k$ centered Taub-NUT and we would have to include the appropriate $B$-fluxes.

\section{Weak coupling limit and the $\cS_k$ free trinion}
\label{sec:Trinion}

The free trinion $\mathcal{C}_{0,3}^{(k)}$ is a
basic building block for class $\cS_k$ theories and can be obtained from the weak coupling
limit of the $SU(N)$ SCQCD$_k$ theory, the four-punctured sphere $\mathcal{C}_{0,4}^{(k)}$. For simplicity, but without any loss of physics information, in this section we will present
in detail only the $N=2$ case. The results generalize to any $N$ immediately.

\subsection{The curves and their punctures}

At weak coupling the SW curve \eqref{N=2kCascurve} becomes
\be \label{SWcurveFreeT1}
(v^k-m_1^k)(v^k-m_2^k)t+(-v^{2k}+ u_k\, v^k + u_{2k}) = 0
\ee
by sending $q\rightarrow 0$ and removing the solution where $t$ vanishes identically.
In the limit $t\rightarrow \infty$, the term that dominates is
\be
(v^k-m_1^k)(v^k-m_2^k)t = 0  \quad \Rightarrow \quad v_\pm^{(n)}~_{t\to\infty}=\{m_1^{(n)}, ~ m_2^{(n)}\}
~,\quad n=1,\ldots, k ~.
\ee
This maximal puncture   is precisely identical to that in the previous section for the four-punctured sphere $\mathcal{C}_{0,4}^{(k)}$.
At the second maximal puncture, where $t= 0$, \eqref{SWcurveFreeT1} reduces to
\be \label{solutionv0}
-v^{2k}+ u_k\, v^k + u_{2k}=0 \quad \Rightarrow \quad v_\pm^k=\frac{1}{2}\left[u_k\pm\left(u_k^2+4u_{2k}\right)^{1/2}\right] ~.
\ee
From section \ref{ABintegrals}, the weak coupling relations  \eqref{uka1a2} between the Coulomb moduli imply
\be
u_k=a_1^k+a_2^k ~, ~ u_{2k}=-a_1^ka_2^k \quad \Rightarrow \quad u_k^2+4u_{2k} =
(a_1^k-a_2^k)^2 ~,
\ee
so the solutions \eqref{solutionv0} can be rewritten
\be \label{Trisolutionatt0}
v_\pm^{(n)}~_{t\to 0}=\{a_1^{(n)}, ~a_2^{(n)}\} ~.
\ee
The $SU(2)$ gauge groups in the weak coupling limit become ungauged flavor symmetry groups and the Coulomb moduli should be replaced by mass parameters $a_1\rightarrow m_3,~ a_2\rightarrow m_4$. This replacement
brings the SW curve in equation \eqref{SWcurveFreeT1} to the form
\be
 \label{SWcurveFreeT2}
 \boxed{
\prod_{i=1}^2 (v^k-m_i{}^k)t + \prod_{i=3}^{4} (v^k-m_i{}^k) = 0~
} ~
\ee
which is identified as a free trinion theory.
This is the spectral curve of four orbifolded free hypermultiplets, with solutions $v^k$ at general $t$
\be
\label{solFtrinionV}
v_\pm^k=\frac{\fc_L^{(1,k)}t-\fc_R^{(1,k)}\pm\sqrt{\Delta}}{2(t-1)}~,\quad
\Delta=(\fc_L^{(1,k)}t-\fc_R^{(1,k)})^2-4(t-1) (\fc_L^{(2,k)}t-\fc_R^{(2,k)})~.
\ee
The UV curve described by this equation is a three-punctured sphere $\cC_{0,3}^{(k)}$. To see this,
we first rewrite the spectral curve \eqref{SWcurveFreeT2} in the form
\be
\label{gaiottoexample1}
(t-1)v^{2k} - (\fc_{L}^{(1,k)} t - \fc_{R}^{(1,k)}) v^k + ( \fc_{L}^{(2,k)} t -
\fc_{R}^{(2,k)})  = 0 ~.~~
\ee
Then substituting $v=xt$ and reparametrizing $t\rightarrow\frac{az+b}{cz+d}$ and $x\rightarrow(cz+d)^2x$
like in equation \eqref{Neq3si2NS5b2}  brings the curve into the form
\be \label{curveandsphere3}
x^{2k}=\frac{\fp_1(z)}{(cz+d)^k(\alpha z+\beta)(az+b)^k} x^k + \frac{\fp_1'(z)}{(cz+d)^{2k}
(\alpha z+\beta)(az+b)^{2k}} ~,
\ee
where $\fp_1$ and $\fp_1'$ are degree one polynomials of $z$. Thus $z$ parametrizes the space $\cC_{0,3}^{(k)}$.

The position $z=-\beta/\alpha$, which corresponds to $t=1$, is that of a minimal puncture. Here, the
solution $v_+^k$ of equation \eqref{solFtrinionV} is singular whereas $v_-^k$ is regular
\be \label{solFtrinionVt1}
v_+^k~_{(t=1)}\sim \frac{1}{t-1}(\fc_L^{(1,k)}-\fc_R^{(1,k)})-
\frac{\fc_L^{(2,k)}-\fc_R^{(2,k)}}{\fc_L^{(1,k)}-\fc_R^{(1,k)}} ~, \qquad v_-^k~_{(t=1)}\sim
\frac{\fc_L^{(k,2)}-\fc_R^{(k,2)}}{\fc_L^{(k,1)}-\fc_R^{(k,1)}} ~.
\ee
This minimal puncture 
does not correspond to a simple pole of the SW differential \eqref{SWdifferential}, but to a branch
point with fractional power $v_+ \sim \left(t-1 \right)^{-1/k}$.
Moreover, the integral around a closed cycle is zero. This puncture is therefore the same as the
minimal punctures on $\mathcal{C}_{0,4}^{(k)}$.

\subsection{Cut structure}
\label{sec:cutsTrinion}

%
\begin{figure}[ht]
\centering
\includegraphics[width=0.7\textwidth]{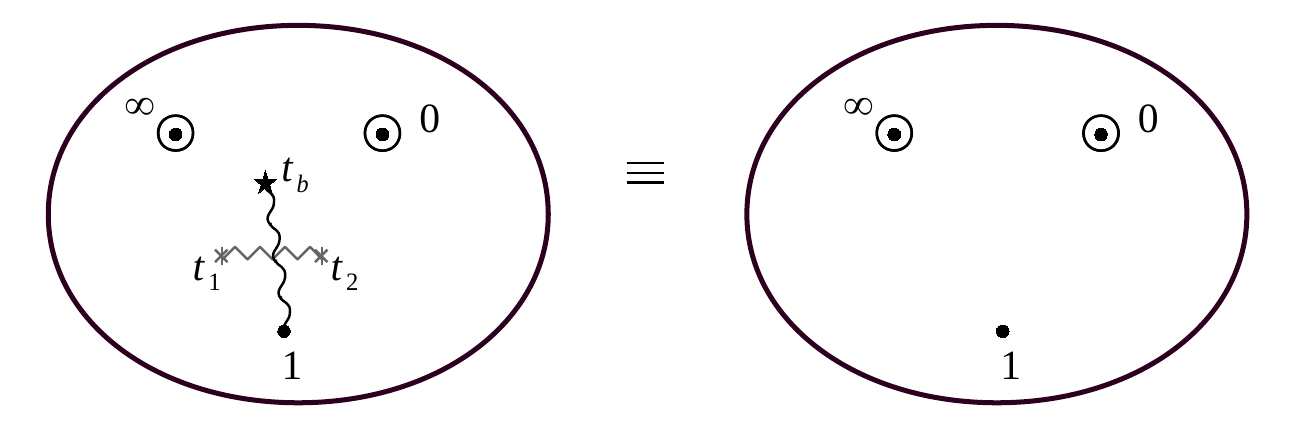}
\caption{\it On the left we draw the IR curve $\Sigma$ with its  branch cut structure. On the right side we depict  the three-punctured sphere $\mathcal{C}_{0,3}^{(k)}$.}
\label{2FreeTrinion1}
\end{figure}
We now turn to the structure of the branch cuts for the free trinion.

\paragraph{Cuts inherited from the $\cN=2$ theory:}
There are two branch points $t_{1,2}$, where the $v_\pm(t_i)$ solutions are equal and which are solutions of the ``discriminant''
\be
\label{N=2discriminant}
\Delta (t_{i}) = (\fc_L^{(1,k)}t_{i}-\fc_R^{(1,k)})^2-4(t_{i}-1)
(\fc_L^{(2,k)}t_{i}-\fc_R^{(2,k)}) = 0~, \quad i\in\{1,2\} ~.
\ee
These points are 
\bea
t_{1,2} = \frac{\fc_L^{(1,k)}\fc_R^{(1,k)} - 2(\fc_L^{(2,k)}+\fc_R^{(2,k)}) \pm
2\sqrt{\Delta_m}}{(m_1^k-m_2^k)^2} ~,
\eea
with
\be
\Delta_m=(m_1^k-m_3^k)(m_2^k-m_3^k)(m_1^k-m_4^k)(m_2^k-m_4^k) ~.
\ee
They are joined by a branch cut $(t_1,t_2)$ in figure \ref{2FreeTrinion1}, which connects the $v_\pm$
sheets.  We can recover  this cut 
in the weak coupling limit of  the cut structure on 
$\mathcal{C}_{0,4}^{(k)}$ in
figure \ref{fig:Gaiotto}. Recall that for the $\mathcal{N}=2$ $SU(2)$ theory
the $A$-cycle shrinks at weak coupling\footnote{This can be easily seen from the massless limit.}. The cut $(t_3,t_4)$  in
figure \ref{fig:Gaiotto}
therefore shrinks at $t=0$, leaving only $(t_1,t_2)$ in figure \ref{2FreeTrinion1}.
Imposing the orbifold
$v_\pm\rightarrow v_\pm^k$ generates $k$ solutions, 
so $(t_1,t_2)$ connects pairwise the sheets $(v_+^{(n)},~v_-^{(n)})_{n=1,\ldots , k}$.

\paragraph{Novel cuts due to the orbifold:}
The orbifold introduces a new branch cut. Equations \eqref{solFtrinionVt1} show that
$v^k_+ ~_{(t\rightarrow 1)}$ diverges and therefore $t=1$ is a branching point. 
This is connected to the point $t=t_b$
\be
t_b=\frac{m_3^km_4^k}{m_1^km_2^k} =\frac{ \fc_R^{(2,k)} }{\fc_L^{(2,k)} } ~,
\ee
where the solution $v^k_-$ vanishes while $v_+^k$ remains
finite
\bea
v_-^k~_{(t \to t_b)} \sim  \frac{\fc_L^{(2,k)}~ (t-t_b)}{\fc_L^{(1,k)}t_b-\fc_R^{(1,k)}} ~,
\quad
v_+^k~_{(t \to t_b)} \sim \frac{\fc_L^{(1,k)}t_b-\fc_R^{(1,k)}}{t_b-1}~.
\eea
The locus on the $t$-space which corresponds to
this cut can also be determined using the knowledge that we have obtained the free trinion
in the weak coupling limit of the curve \eqref{N=2kCascurve}. The value $t_b$ in equation
\eqref{VminusBP} goes to $ \fc_R^{(2,k)} / \fc_L^{(2,k)}$, whereas $\tilde t_b \to 0$.

\section{Closing a minimal puncture and the free trinion theory}
\label{Closing}

Trinions are the fundamental building blocks of class $\cS_k$ theories and it is therefore important
to investigate the different ways to construct them. In the previous section we have found
the free trinion in the weak coupling limit of the four-punctured sphere $\cC_{0,4}^{(k)}$.
In this section, 
we begin with the curve for $SU(N)$ SCQCD$_k$ \eqref{SWcurveNS52Zk},
which has two maximal punctures at $t=0, \infty$ and two minimal punctures at $t=q,1$, and
close the puncture at $t=q$.  We compute the corresponding spectral curve of the resulting trinion.

 Using the superconformal index as an avatar, Gaiotto and Razamat \cite{Gaiotto:2015usa} found an interacting trinion theory  by closing a minimal puncture.
 From the point of view of the index, this interacting trinion is different  from the free trinion obtained at  weak coupling.
The difference between the two trinion theories is due to an new quantum number, that they introduce as ``color''.
The ``color'' is related to shifting the labels of the $U(1)_{\beta_i}$ and/or $U(1)_{\gamma_i}$ in table \ref{tab:SymmetriesSCQCD}.
In this paper we study the curves on the Coulomb branch and turn on only glueball type operators
$\mbox{tr} \left( \Phi_{(1,c)} \cdots \Phi_{(k,c)} \right)^\ell$,
which are not charged under $U(1)_\beta$  and $U(1)_\gamma$ and thus do not transform  under the ``color''.
To feel a difference in color, one should give vevs to mesons and go to the Higgs branch.
In this paper we stay on the Coulomb branch, without turning on vevs for the mesons, and present below the procedure of closing a minimal puncture.
The trinion which we find results to have the same structure as the free trinion in section \ref{sec:Trinion}.

\begin{figure}[h!]
\centering
\includegraphics[width=0.4\textwidth]{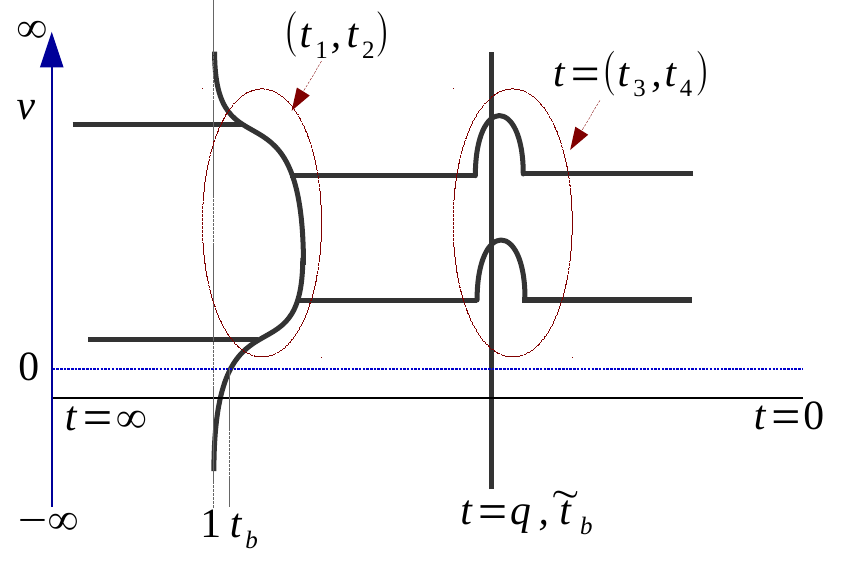}
\caption{\it Closing a minimal puncture in the type IIA setup for $SU(2)$ SCQCD.}
\label{IIAbraneclose}
\end{figure}

\paragraph{Closing a puncture:}
The easiest way to understand the closure of a minimal puncture is to consider the type IIA setup depicted in figure \ref{IIAbraneclose}.
This procedure corresponds to tuning the position of the right flavor and color branes to be identical.
They behave as if they penetrate the NS5 brane on the right, whereby this becomes straight at the position $t=q$.

At the level of the curve \eqref{SWcurveNS52Zk}, this is realized by properly tuning the mass parameters and Coulomb moduli
in such a way that $v(t)$ does not have a singularity at $t=q$.
Therefore, we need to impose $P_n(t=q) = 0$, which leads to
\bea
u_{nk} = (-1)^{n-1} ( \fc_L^{(n,k)}q + \fc_R^{(n,k)} )~.
\label{unk}
\eea
Under this tuning of parameters, the curve reduces to
\be
\label{SWcurveTrinionFreeN}
\prod_{i=1}^N (v^k-m_i{}^k)t + \prod_{i=N+1}^{2N} (v^k-m_i{}^k) = 0~
\ee
after factoring out $(t-q)$.
This is indeed identical to the curve for the free trinion \eqref{SWcurveFreeT2} depicted in figure \ref{2FreeTrinion1}.
Below we describe how the branching structure from figure \ref{fig:Gaiotto} is modified when closing a minimal puncture.


\subsection{Cut structure}
\label{sec:cutstrMP}

Here, we study the structure of the branch cuts modified by closing the puncture $t=q$ of the
sphere $\cC_{0,4}^{(k)}$. We consider $N=2$ for simplicity.
The cuts in figure \ref{fig:Gaiotto} are modified by tuning parameters as in \eqref{unk}.
To see this process, we first set
\be \label{conditionsclosepunctMgen}
u_k=\fc_R^{(1,k)}+q\fc_L^{(1,k)} ~,\quad
u_{2k}=-\fc_R^{(2,k)}-q\fc_L^{(2,k)} ~.
\ee

\paragraph{Cuts inherited from the $\cN=2$ theory:} First, we consider the four branch points $t_i$
which are the solutions of $\Delta=0$ defined in \eqref{quadraticdiscriminant}.
This factorizes as
\be
\Delta = (t-q)^2
\left( (t \fc_L^{(1,k)} - \fc_R^{(1,k)})^2 - 4 (t-1)(t \fc_L^{(2,k)}- \fc_R^{(2,k)}) \right) ~,
\label{CloseDelta}
\ee
thus two of the branch points approach the position $t=q$.
To be consistent with the convention in appendix \ref{app:B},
we identify $t_3$ and $t_4$ to be these points.
Note that $t=q$ is exactly the point where the closed minimal puncture was placed.
Since there is no puncture at $t=q$ after tuning the Coulomb moduli,
both the cut $(t_3,t_4)$ and the puncture $t=q$ vanish
by mutually cancelling their effect. The cut $(t_1,t_2)$ however remains.

\paragraph{Novel cuts due to the orbifold:}
Next, we consider the cuts between ($t_\bullet,  t_\bigstar$).
The values $t_b$ and $\tilde t_b$ from equation \eqref{VminusBP} go to
\be
t_b = \frac{\fc_R^{(2,k)}}{\fc_L^{(2,k)}} ~, \qquad
\tilde t_b = q ~.
\ee
Here again, we see that $\tilde t_b$ approaches $t \to q$.
Thus the branching points $t=q$ and $t=\tilde t_b$ collide and the branch cut that connects them
disappears. Only the cut between $t=1$ and $t=t_b = \fc_R^{(2,k)}/\fc_L^{(2,k)}$ remains, which
is the same as for the free trinion depicted in figure \ref{2FreeTrinion1}.

\section{Curves with $M>2$ minimal punctures}
\label{M>2minimal}

In this section we wish to study theories whose curve is the $(M+2)$-punctured sphere, with two maximal and $M$  minimal punctures.
To do so, as reviewed in section \ref{sec:Setup}, we begin with the conformal $\cN=2$ linear quivers with  $SU(N)^{M-1}$ gauge groups and superpotential \eqref{eq:N=2superpotential}
 and then impose
the $\bZ_k$ orbifold. In the type IIA set-up these are obtained by having   $M$
NS5 branes on which the $N(M-1)$ D4 color branes end.
We use the $\cN=2$ SW curves computed in \cite{Bao:2011rc} and then orbifold by
 imposing the identifications \eqref{orbiPrefRep}. The positions of the $M$
NS5 branes along the $x^6$ direction are parametrized by $q_\alpha$, with $\alpha=1,\ldots,M-1$.

The resulting curves after the orbifold are
\be
\label{generalNS5b}
\prod_{i=1}^N(v^k-m_i^k)t^M +
\sum_{\alpha=1}^{M-1} {\mathfrak{P}}_\alpha(v)  \,t^{M-\alpha}+
(-1)^M \prod_{\alpha=1}^{M-1}q_\alpha^{M-\alpha}\prod_{i=N+1}^{2N}(v^k-m_i^k)=0 ~ ,
\ee
where
\be
{\mathfrak{P}}_\alpha(v)=d_\alpha v^{Nk} + u_{k,\alpha} v^{(N-1)k} +\ldots + u_{Nk,\alpha}
\ee
and the coefficients $d_\alpha$ are
\be
d_\beta=(-1)^\beta \sum_{1\leq \alpha_1<\alpha_2<\ldots <\alpha_\beta\leq M} t_{\alpha_1}t_{\alpha_2}\ldots t_{\alpha_\beta} \quad \text{for}
\quad q_\alpha=\frac{t_{\alpha+1}}{t_{\alpha}}~, ~ t_{\alpha =1}=1 ~.
\ee
In terms of the parameters  $q_\alpha$, $d_1=-(1+\sum_{\alpha=2}^M\prod_{\beta=1}^{\alpha-1}q_\beta)$.
Equation \eqref{generalNS5b} is equivalent to
\be \label{eq:genNM}
\prod_{\alpha=1}^{M}(t-t_\alpha)v^{Nk} = P_1(t) v^{(N-1)k} - P_2(t) v^{(N-2)k} + \ldots + (-1)^{N-1} P_N(t) ~.
\ee
The $M$ values $t_\alpha$ are $t_{\alpha=1}=1$ and
$t_{\alpha>1}=\prod_{\beta=1}^{\alpha-1}q_\beta$ and $P_i(t)$ are polynomials of order $M$.

The SW differential at the maximal punctures $t=0,\infty$ behaves in the same way as for the
curves studied in the previous sections. It has simple poles on all of the $Nk$ sheets
described by the solutions to equation \eqref{eq:genNM}, with residues $m_i^{(n)}$.
To gain insight about the minimal punctures and the cut structure of this curve, it is
helpful to begin by setting $N=2$. The generalization is then immediate.

\begin{figure}[t]
 \begin{center}
\includegraphics[width=10cm]{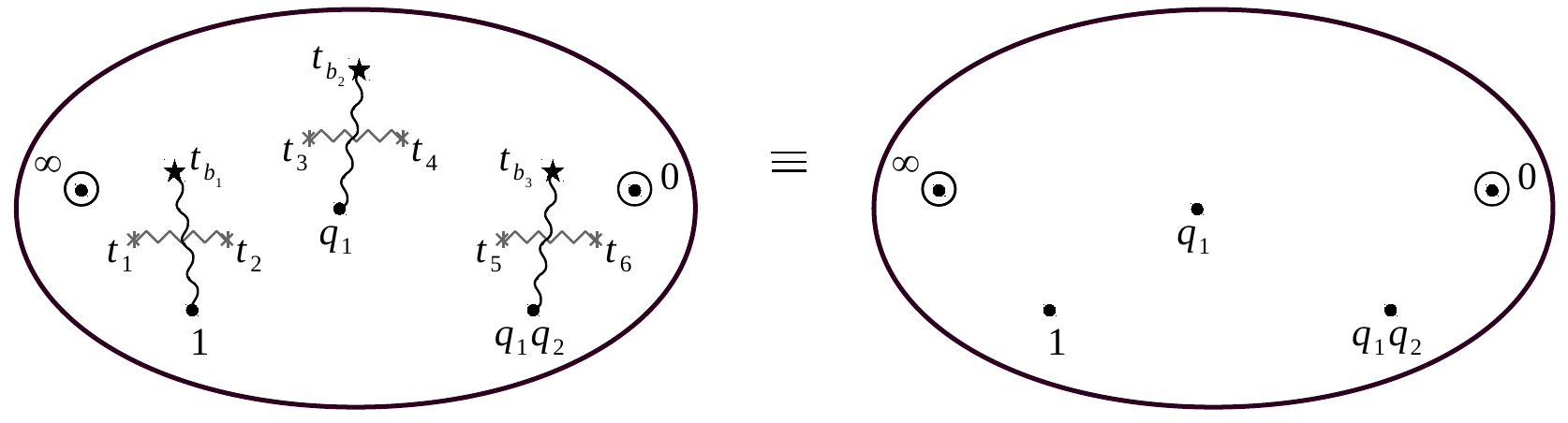}
 \end{center}
 \caption{ \it The IR spectral curve of the orbifolded $SU(2)^2$ quiver obtained as a $2k^{th}$ cover of the five-punctured sphere in class $\cS_k$. On the left we draw the IR curve $\Sigma$ and explicitly denote the branch cut structure. On the right side depict only the five-punctured sphere $\mathcal{C}_{0,5}^{(k)}$.}
 \label{fig:5puncturedSphere}
\end{figure}

\subsection{The case $N=2$}

The polynomials $P_i$ in equation \eqref{eq:genNM} are
\bea
P_1(t) &=&
\fc_L^{(1,k)}t^M - u_{k,1}t^{M-1}+\ldots -u_{k,M-1}t
+(-1)^M \prod_{\alpha=1}^{M-1}q_\alpha^{M-\alpha}\fc_R^{(1,k)} ~,  \\
P_2(t) &=&
\fc_L^{(2,k)}t^M + u_{2k,1}t^{M-1}+\ldots +u_{2k,M-1} t
+(-1)^{M} \prod_{\alpha=1}^{M-1}q_\alpha^{M-\alpha}\fc_R^{(2,k)} ~.
\eea
The plus-type solutions from the set $v_\pm^k$
\be
\label{genk2solvk2}
v^k_\pm = \frac{P_1(t) \pm \sqrt{P_1(t)^2-4\prod_{\alpha=1}^{M}(t-t_\alpha)P_2(t)}}
{2\prod_{\alpha=1}^{M}(t-t_\alpha)} ~
\ee
develop a pole as $t\to t_\alpha$
\be
v_+^k(t_\alpha) \sim \frac{P_1(t_\alpha)}{(t-t_\alpha) \prod_{\alpha\neq\beta}(t_\alpha-t_\beta)} ~.
\ee
The points $t= t_\alpha$ are the positions of the minimal punctures, similar to those of the
$M=2$ cases studied previously. The solutions $v_-^k(t_\alpha)\sim P_1(t_\alpha)/P_2(t_\alpha)$ remain regular at these points.

\paragraph{Branch cuts inherited from $\cN=2$:} The curve described by equation \eqref{eq:genNM} has
branch points inherited from the $\cN=2$ theory at the values $t=t_i$, where
\be
P_1(t_i)^2-4\prod_{\alpha=1}^{M}(t_i-t_\alpha)P_2(t_i) = 0 ~, \quad i=1,\ldots , 2M ~.
\ee
These points create $M$ branch cuts that connect the sheets described by the $v_\pm^k$ solutions.

\paragraph{Novel branch cuts:} In addition to the $\cN=2$ inherited cuts, the orbifold introduces novel ones.
These connect pairwise the branching points
$t=t_\alpha$ where $v^k_+$ solutions develop a singularity, to $M$ branching points $t=t_{b_\alpha}$
where $v^k_-$ solutions vanish. The values $t_{b_\alpha}$ where this happens are precisely the zeros
of the polynomial $P_2(t)$. Therefore the curve described by equation \eqref{eq:genNM} has genus $(M-1)(2k-1)$.

\begin{figure}[t]
\centering
\includegraphics[width=0.5\textwidth]{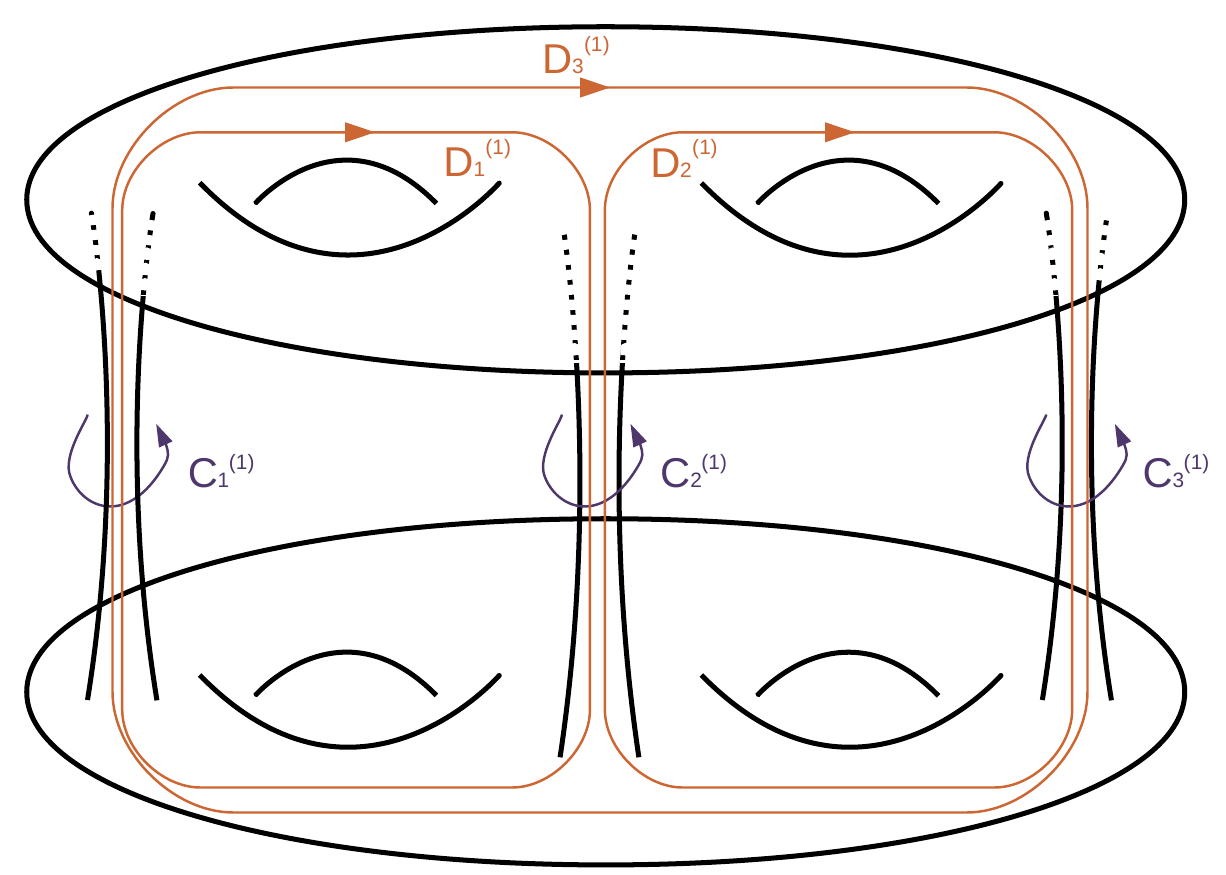}
\caption{\it The IR curve $\Sigma$ for the $SU(2)^2$ quiver with $k=2$ and its novel cycles, due to the $\bZ_2$ orbifold. We have taken the massless limit where the maximal punctures disappear.}
\label{fig:Z2-2Torus}
\end{figure}

Due to these novel branch cuts, we have new cycles
which generalize the
$C_i^{(n)}$ and $D_i^{(n)}$ in section \ref{ABintegrals}.
The case with two maximal punctures and three minimal punctures
for $N=k=2$ is depicted in Figure \ref{fig:Z2-2Torus}.
The $C$-cycle integrals correspond to the center of mass of the color
D4-branes while the $D$-cycles are related  non-trivially to the other cycles like in \eqref{DcyclesIntegrals}.

\subsection{General $N$}

One can easily generalize these results to any $N$.
Equation \eqref{eq:genNM} for the spectral curve has  $N$ solutions $v^k_{(i)}$,
which are connected by $(N-1)M$ branch cuts inherited from the mother $\cN=2$ theory.
One of these solutions diverges at all minimal punctures $t=t_\alpha$, which are therefore branching
points with fractional singularities.
There exist furthermore $M$ solutions $v^k$ which vanish at the values $t=t_{b_\alpha}$
that are the zeroes of the polynomial $P_N(t)$. The positions $t=t_\alpha$ and
$t=t_{b_\alpha}$ are thus branching points connected by $M$ branch cuts and the
genus of the curve \eqref{eq:genNM} is $(M-1)(Nk-1)$.

\bigskip

We would like to conclude this section by counting the total number of parameters that appear in our curves. The result of the counting is summarized in table  \ref{tab:GcurveParameres}, where the total number of parameters the curves of the $\cN=1$ class $\cS_k$ is compared with those in the curves of class $\cS$.  The following comments are in order. When we count mass parameters and Coulomb moduli in class $\cS_k$, we take into account that they satisfy the orbifold condition and get identified as \eqref{orbifoldrestriction1}
\be
m_i^{(n)}=  e^{\frac{2 \pi i n}{k}} m_i
~, \quad
a_i^{(n)}=  e^{\frac{2 \pi i n}{k}} a_i
~,\qquad n=1,\ldots,k
~.
\ee
This means that for each stack of $N$ D4 branes and their mirror images we only have $N$ independent mass parameters in our curves  and each full puncture has $U(N)$ symmetry.
This is in contradistinction with the $\cN=2$ case. As we already stressed in section \ref{sec:curves},  in the case where we have an $\cN=2$ theory with a single color group,  we where allowed to shift the origin of $v$ so that it coincides with the center of mass of the color D4 branes with the consequences: $(i)$
 $\sum_{i=1}^Na_i =0$, $(ii)$
 the mass parameters of a maximal puncture transform in the $SU(N)$ instead of the $U(N)$ and  $(iii)$  the $U(1)$ went from the maximal puncture to the minimal puncture.
In the case where we have more color groups we are not allowed any more to set $\sum_{i=1}^Na_i =0$ for every color group. However, these trace combinations  give the bifundamental masses of the theory
\be
m_{bif}^{(\alpha)} \sim \sum_{i=1}^Na_i^{(\alpha)}  - \sum_{i=1}^Na_i^{(\alpha-1)}  \quad \mbox{with} \quad \alpha=2,\dots,M ~.
\ee
For $\cN=2$  theories,  the minimal punctures have $U(1)$ symmetry  parameterized by these bifundamental masses.
When we then turn to the $\cN=1$ theories of class $\cS_k$, we can no longer turn on bifundamental masses. There is no gauge invariant term for   bifundamental masses that could be added to the Lagrangian. However, the combinations $ \sum_{i=1}^Na_i^{k}$ are still there and non-zero,  they just have a different place in the curve. They are obtained when performing the novel, $C$-cycle integrals. An example was explicitly calculated  in
\eqref{CcycleIntegral}.

\begin{table}[t]
 \centering
    \begin{tabular}{|c || c | c  |  c | c| c|}
    \hline
  &    masses $m_i$ & Coulomb moduli $u_{i,j}$ & $q_j$ & bif. masses & total
   \\
    \hline
	 \hline
class $\cS$ &	$2N$  &  $(M-1)(N-1)$  &  $M-1$   &  $M-2$      &   $MN +N+ M-2$
	 \\
	  \hline
class $\cS_k$  &	$2N$  &  $(M-1)N$  &  $M-1$      &  $0$   & $MN +N+ M-1$
	 \\
	 \hline
    \end{tabular}
	 \caption{\it  Counting the parameters that appear in the $\cN=2$ class $\cS$ and the $\cN=1$ class $\cS_k$ spectral curves, for $M+2$ punctured spheres.
	 Notice that class $\cS_k$ has one more parameter in total. This is related to the fact that we are not allowed to shift $v$ as the origin is set at the  orbifold fixed point.}
	 \label{tab:GcurveParameres}
\end{table}


\section{Strong coupling and the non-Lagrangian $T_N^k$ theories}
\label{TNk}

In this section, we study orbifolding the $T_N$ theories.
The $T_N$ theories are some of the most typical non-Lagrangian theories in class $\mathcal{S}$
and are described by $N$ M5-branes wrapping a sphere with three maximal punctures.
Their SW curves, which we use here, were computed in \cite{Bao:2013pwa}. See also \cite{Benini:2009gi}.
Orbifolding means that we impose the identification $v\sim e^{\frac{2\pi i}{k}}v$, and we denote the theories obtained by $T_N^k$.
In class $\mathcal{S}$, the $T_N$ theories are obtained by taking the  strong coupling limit of $SU(N)^{N-2}$ linear superconformal quivers.
Similarly the $T_N^k$ theories are obtained by taking the  strong coupling limit of $SU(N)^{N-2}_k$ quivers. We demonstrate this point explicitly for $SU(N)$ SCQCD$_k$
for $N=2,3$. 

It is important to stress that among the three maximal punctures at $t=0,1,\infty$,
 the $\mathbb{Z}_k$ orbifold acts only on the two labelled by $\odot$, at $t=0,\infty$. These punctures correspond to simple poles of $\lambda_{SW}$, while the $t=1$ puncture labelled by $\StarPt$ corresponds to a fractional singularity.\footnote{We wish to note that also \cite{Gaiotto:2015usa} find that two out of three maximal punctures of the $T_N^k$ theories are the same while one is different  using the superconformal index.   From the point of view of the index  the difference lies on the different ``color'' of the punctures.}
We cannot help but  wonder if there are theories in the class $\cS_k$ that correspond to three punctured spheres with either all orbifolded $\odot$ simple poles punctures, or all unorbifolded $\StarPt$ fractional singularities.

\begin{table}[h!]
\centering{\renewcommand{\arraystretch}{1.5}
\begin{tabular}{|c|c|c|}
\hline
$\odot$ & Maximal puncture &  $(t-t_\odot)^{-1}$  \\
\hline
$\StarPt$ & Fractional maximal  &  $(t-t_{\StarPts})^{-\frac{1}{k}}$\\
\hline
\end{tabular}   }
\caption{\it The two different kinds of maximal punctures on the $T_N^k$ curve.}
\label{TN-legend-points}
\end{table}

\subsection{The curves}

We begin with the SW curve for the $T_N$ theory obtained in \cite{Bao:2013pwa}, with $N=2,3$.
The SW curve of the $T_2$ theory is given by
\bea
(v-m_1) (v-m_2)t^2 +
\left( -2v^{2}+\, M_{(1)}^{T_2} \,  v - M_{(2)}^{T_2} \right) t
+ (v-m_3)(v-m_4) &=& 0 ~,
\\
\mbox{with}  \qquad  M_{(1)}^{T_2}  = \sum_{i=1}^4 m_i
 \qquad
M_{(2)}^{T_2} =  \sum_{i < j} m_i m_{j} \, .
\eea
This is the curve of four free hypermultiplets
which is also obtained as the strong coupling limit $q \to 1$ of the SW curve for $SU(2)$ with $N_f=4$ flavor.
The SW curve of the $T_3$ theory is identified as the curve with $E_6$ global symmetry \cite{Minahan:1996fg} and
is given by
\bea
&& (v-m_1)(v-m_2)(v-m_3) t^2
+ \left(
- 2 v^3 +  M_{(1)}^{T_3}  \,v^2
-M_{(2)}^{T_3} \, v
- u
\right) t
\nn \\
&& + (v-m_4)(v-m_5)(v-m_6)  = 0
\\
&& \mbox{with}  \qquad  M_{(1)}^{T_3}  = \sum_{i=1}^6 m_i
 \qquad
M_{(2)}^{T_3} =  \sum_{i < j} m_i m_{j}
\eea
This curve is obtained in the strong coupling limit $q \to 1$ of the
SW curve for $SU(3)$ with $N_f=6$ flavor,
which is understood as Argyres-Seiberg duality \cite{Argyres:2007cn}.

Now, we propose that orbifolding the curve amounts to replacing
\be
(v-m_i) \longrightarrow (v^k-m_i^k) =  \prod^{k}_{n=1} (v-m_i^{(n)})
\ee
and
\be
M_{(1)}^{T_N^k}  = \sum_{i=1}^{2N} m_i^k
 \qquad
M_{(2)}^{T_N^k} =  \sum_{1\le i < j \le 2N} m_i^k m_{j}^k
\qquad (N=2,3) ~.
\ee
The idea here is that the orbifold acts on the $v$ coordinate.
Thus we obtain
\bea
(v^k-m_1^k) (v^k-m_2^k)t^2 +
\left( -2v^{2k}+ \, M_{(1)}^{T_2^k} \,  v^k - M_{(2)}^{T_2^k} \right) t
+ (v^k-m_3^k)(v^k-m_4^k) &=& 0 ~,
\eea
for $T_2^k$, while for $T_3^k$ we find
\bea
(v^k-m_1^k)(v^k-m_2^k)(v^k-m_3^k) t^2
+ \left(
- 2 v^{3k} +  M_{(1)}^{T_3^k}  \,v^{2k}
-M_{(2)}^{T_3^k} \, v^k
- u_{3k}
\right) t
\nonumber \\
+ (v^k-m_4^k)(v^k-m_5^k)(v^k-m_6^k)
= 0 ~.
\eea
These results are interpreted in terms of the orbifolded version of Argyres-Seiberg duality.
It is straightforward to check that the spectral curves for $T_2^k$ and $T_3^k$
are obtained from those of SCQCD$_k$ for $N=2$ and $N=3$ in the strong coupling limit.

Now with an analysis analogous to the one in section \ref{sec:punctures} we look for the punctures of the curve. Around $t=1$ the curve behaves as
\be
\left. v_{\pm}{}\right| _{t\rightarrow 1} \sim \left\lbrace
\frac{(m^k_1+m^k_2)^{1/k}}{(t-1)^{\frac{1}{k}}}~,~ \frac{(-m^k_3-m^k_4)^{1/k}}{(t-1)^{\frac{1}{k}}}\right\rbrace
\ee
for $N=2$. We thus have a new type of puncture with a pole with fractional power that gives us  branch points.
At $t \to 0$ and $t \to \infty$, the curve behaves as
\be
\displaystyle \lim_{t\rightarrow 0}
~v_\pm^{(n)}=
\left\lbrace
m_3^{(n)}, ~ m_4^{(n)}
\right\rbrace_{n=1,\ldots ,k} ~,
\qquad
\displaystyle \lim_{t\rightarrow\infty}
~v_\pm^{(n)}=
\left\lbrace
m_1^{(n)}, ~ m_2^{(n)}
\right\rbrace_{n=1,\ldots ,k} ~
\ee
which are the ``orbifolded'' $U(2)$ maximal punctures encountered in the previous sections and parameterized by $k$ mirror images of $U(2)$.

\begin{figure}
\centering
\includegraphics[width=0.4\textwidth]{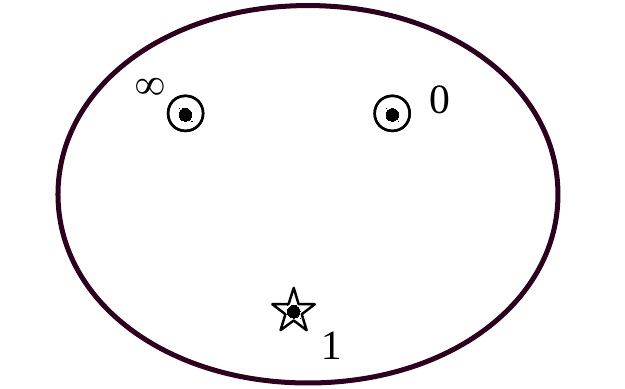}
\caption{\it The spectral curve of the trinion theory $T_N^k$. As opposed to the $\cN=2$ case, here we have two different kinds of maximal punctures.}
\label{fig:TNk}
\end{figure}

The cases with $N\geq 3$ can be studied analogously.
They are obtained as the strong coupling limit of $SU(N)^{N-2}_k$ quivers.
When $N=3$, the curve behaves as
\be
\label{minimalTNk}
\left. v_{i=1,2,3}{}\right| _{t\rightarrow 1} \sim \left\lbrace
\mbox{finite}  ~,~
\frac{(- \fc_R^{(1,k)} )^{1/k}}{(t-1)^{\frac{1}{k}}}~,~ \frac{( \fc_L^{(1,k)}  )^{1/k}}{(t-1)^{\frac{1}{k}}}\right\rbrace \,
\ee
around $t=1$, with
\be
\label{cLcRT3}
\fc_R^{(1,k)}  = m^k_4+ m^k_5+ m^k_6   \quad  \mbox{and}  \quad    \fc_L^{(1,k)} = m^k_1+ m^k_2 + m^k_3 \, .
\ee
 We thus have a new type of maximal puncture with a pole with fractional power that gives us  branch points.
At $t \to 0$ and $t \to \infty$, the curve behaves as
\be
\displaystyle \lim_{t\rightarrow 0}
~v_{i=1\dots N}^{(n)}=
\left\lbrace
m_{N+1}^{(n)}, \dots ,  m_{2N}^{(n)}
\right\rbrace_{n=1,\ldots ,k} ~,
\quad
\displaystyle \lim_{t\rightarrow\infty}
~v_{i=1\dots N}^{(n)}=
\left\lbrace
m_1^{(n)}, \dots ,  m_N^{(n)}
\right\rbrace_{n=1,\ldots ,k} ~.
\ee
We summarize all the possible distinguished points in table \ref{TN-legend-points} and depict them on the UV curve in  figure \ref{fig:TNk}.

\subsection{Cut structure}

For simplicity we begin with the cut structure of the spectral curve for $N = 2$.
In the strong coupling limit $q \to 1$ the discriminant \eqref{quadraticdiscriminant} factorizes as
\be
\Delta = (t-1)^2 \left( (\fc_L^{(1,k)} t-\fc_R^{(1,k)} )^2 - 4( \fc_L^{(2,k)}t^2 - M_2^{T_2^k} t+\fc_R^{(2,k)} ) \right) ,
\ee
which means that two of the four branch points approach the position $t = 1$.
From the convention (\ref{masslessconvention}),
we identify  these two points to be $t_2$ and  $t_3$.
The $B$-cycle thus shrinks and the branch cuts  $(t_1,t_2)$ and  $(t_3,t_4)$
 combine into one.

Turning to the novel cuts due to the orbifold, on  top of the  $t=1$ branch points that we have found above
\be
\label{t=1BPTN=2}
\left. v_{\pm}{}\right| _{t\rightarrow 1} \sim \left\lbrace
\frac{(m^k_1+m^k_2)^{1/k}}{(t-1)^{\frac{1}{k}}}~,~ \frac{(-m^k_3-m^k_4)^{1/k}}{(t-1)^{\frac{1}{k}}}\right\rbrace
\ee
we also have two branching points where $v\sim (t-t_b)^{1/k}$. These are
\be
\label{tbBPTN=2}
t_b , \tilde{t}_b = \frac{M_{(2)}^{T_2^k} \pm \sqrt{\left(M_{(2)}^{T_2^k}\right)^2-4 m_1^k m_2^k m_3^k m_4^k}}{2 m_1^k m_2^k} \, .
\ee
The $t=1$ branch points at $v_\pm\sim (t-t_b)^{-1/k}$ \eqref{t=1BPTN=2} are therefore connected by branch cuts to $t_b , \tilde{t}_b$,
where $v\sim (t-t_b)^{1/k}$.

The generalization to  $N\geq 3$ can be obtained analogously.
From the branch cuts that are inherited from $\cN=2$, $N(N-1)/2$ remain.
The novel cuts for $N=3$ connect the $t =1$  branch points, where the
 $v_{2}$ and $v_{3}$ solutions diverge \eqref{minimalTNk},  with the $N=3$ version of $t_b , \tilde{t}_b$, that is
\be
t_b , \tilde{t}_b =   \frac{- u_{3k}\pm \sqrt{ u_{3k}^2 -4  \fc_L^{(1,k)}  \fc_R^{(1,k)}   }}{2  \fc_L^{(1,k)}}  \, ,
\ee
where $ \fc_L^{(1,k)}$ and $ \fc_R^{(1,k)}$ are defined in equation \eqref{cLcRT3}.

\section{Conclusions and Outlook}


In this paper we constructed spectral curves for the Gaiotto-Razamat $\mathcal{N}=1$ theories in class $\mathcal{S}_k$  in the Coulomb branch.
We first obtained the IR curve $\Sigma$ for SCQCD$_k$ using Witten's M-theory approach \cite{Witten:1997sc,Witten:1997ep} and then rewrote it  $\grave{a}$ $la$ Gaiotto \cite{Gaiotto:2009we} as a  four-punctured sphere, the UV curve.
By looking at it in  different S-duality frames, we discovered that also for  class $\mathcal{S}_k$  pants decomposition is a valid operation, with trinion theories as the main building blocks  in a spirit similar to the $\mathcal{N}=2$ class $\mathcal{S}$   \cite{Gaiotto:2009we}. The new element is that the  spectral curves have a novel cut structure.
Moreover, the minimal punctures, denoted with $\bullet$, do not  correspond to simple poles, but to branch points with fractional $1/k$ power.
What is more, we discovered two kinds of  maximal punctures. One of them $\odot$ corresponds to simple poles and is labeled by parameters that are $k$ mirror images of $U(N)$. The second one $\StarPt $ is discovered only in the strong coupling limit and corresponds  to a fractional singularity.  We summarize all the different punctures that we discovered in table \ref{All-legend-points-WeCouldFind}.

\medskip

Following Gaiotto and Razamat \cite{Gaiotto:2015usa}
we studied the consequences of  taking the weak coupling limit and
 closing a minimal puncture of the four-punctured sphere $\cC_{0,4}^{(k)}$ at the level of the spectral curve. Both procedures led to the curve of the free trinion theory $\cC_{0,3}^{(k)}$,
 with two maximal $\odot$ and one minimal $\bullet$ punctures.
Finally, we constructed theories that correspond to orbifolding the $T_N$ theories from  class $\mathcal{S}$, which we refer to as $T_N^k$. They can be obtained as the strong coupling limit of $SU(N)^{N-2}_k$ quivers. Two of their maximal punctures $\odot$ are the same as the maximal punctures of $\cC_{0,4}^{(k)}$, while one of them $\StarPt $ is very different.

\begin{table}[t]
\centering{\renewcommand{\arraystretch}{1.5}
\begin{tabular}{|c|c|c|}
\hline
$\odot$ & Maximal puncture &  $(t-t_\odot)^{-1}$  \\
\hline
$\bullet$ &  Minimal puncture &  $(t-t_\bullet)^{-\frac{1}{k}}$\\
\hline
$\bigstar$ &  Branching point $t_b$ &  $(t-t_b)^{\frac{1}{k}}$\\
\hline
$\StarPt $ & Maximal puncture &  $(t-t_{\StarPts} )^{-\frac{1}{k}}$  \\
\hline
\end{tabular}   }
\caption{\it All the different punctures and novel branching points we discovered in the class $\cS_k$.}
\label{All-legend-points-WeCouldFind}
\end{table}

\bigskip

%
%
%
%

 Class $\cS_k$ theories represent a vast topic, which we have only  begun to explore.
 Computing the spectral curves we discovered  novel structures which require further study.
 Furthermore, apart from the spectral curves there exist many other tools which we should employ to investigate class $\cS_k$.

In table \ref{All-legend-points-WeCouldFind} we summarize all the punctures and novel branching points that we discovered in this paper. But can there exist others? Is it possible to have  arbitrary combinations of  these on a Riemann surface or are there restrictions?  The rules of the game need to be understood and a classification is required.
The basic building blocks  of  theories in class $\cS_k$ are trinions. In this paper we discovered the free trinion and the $T_N^k$.
 From table \ref{All-legend-points-WeCouldFind} we can imagine more possibilities.
If they are allowed, we should try to understand them.

\smallskip

The curves that we have computed in this paper are for the theories at the orbifold point. As we showed in section \ref{ABintegrals} all the $k$ YM coupling constants $\tau^{(n)}_{IR}=\tau_{IR}$ for all $n$ are equal to each other. It would be very interesting to generalize our work for theories away from the orbifold point. That would involve replacing the orbifold by a Taub-NUT and further specifying the appropriate periods of $B$-fluxes in the M-theory setup.

A very interesting but simple generalization of our work is to write down also the curves for the more extended class of theories considered in \cite{Franco:2015jna,Hanany:2015pfa}. These theories, from the IIA point of view, allow for an extra rotation by $90^o$ from $v$ to $w$ of some of the NS5 branes and thus the $\mathcal{L}_w$ bundle becomes non-trivial. This is work in progress.

In a very interesting paper \cite{Tachikawa:2011ea} the spectral curve for an $SU(2)^3$ trifundamental  was computed in two different ways. One way was $\grave{a}$ $la$ Gaiotto \cite{Gaiotto:2009we} in which the vevs of the trifundamentals are set
to zero.
The other way was  $\grave{a}$ $la$ Intriligator and Seiberg \cite{Intriligator:1994sm}, where the theory is in a phase where  the trifundamentals acquire a vev. Amazingly, the authors of  \cite{Tachikawa:2011ea} where able to show that the two curves are equivalent, up to some identification of parameters. It would be very interesting to try to do the same for the $\mathcal{N}=1$ theories in the  class $\mathcal{S}_k$. Computing $\cS_k$ curves $\grave{a}$ $la$ Intriligator and Seiberg \cite{Intriligator:1994sm} has the advantage that the vevs for the mesons which are twisted operators (charged under ``color'') under the orbifold immediately appear in the curve.

Apart from the SW curves and the string/M-theory constructions that we discuss in this paper,
 there are further powerful  tools that have revolutionized the study of $\mathcal{N}=2$ theories. These include
 the Dijkgraaf-Vafa matrix models (old and new   \cite{Dijkgraaf:2002fc,Dijkgraaf:2002vw,Dijkgraaf:2002dh,Dijkgraaf:2009pc})
as well as
localization and topological strings.
Studying the Dijkgraaf-Vafa matrix models for  the $\mathcal{N}=1$ theories in class $\mathcal{S}_k$ is a direction worth pursuing  as it allows to  search for relations with CFT's, having an eye on a possible  AGT$_k$ correspondence. This is work in progress.
We also think that it is worth attempting to generalize the work of Pestun for the $\mathcal{N}=1$ theories in class $\mathcal{S}_k$.

Topological strings  have provided an alternative way to obtain the partition function of 5D  $\mathcal{N}=1$ theories on $\mathbb{S}^4\times \mathbb{S}^1$. The partition functions are read off from the corresponding web-toric diagram. To use topological strings for the study of  the $\mathcal{N}=1$ theories in class $\mathcal{S}_k$ we need to uplift to the 5D $\mathcal{S}_k$ theories which are 5D $\mathcal{N}=1$ theories with an extra defect in the 5th dimension. The toric diagram will have to be orbifolded and the partition function in the presence of a defect \cite{Gaiotto:2015una} will have to be computed. See also \cite{Drukker:2010jp}.


\acknowledgments

The authors have greatly benefited from discussions with
Till Bargheer,  Giulio Bonelli, Hirotaka Hayashi, Madalena Lemos, Kazunobu Maruyoshi, Shlomo S. Razamat, Alessandro Tanzini and Joerg Teschner.
F.Y. is thankful to Workshop on Geometric Correspondences of Gauge Theories at SISSA.
The work of IC was supported by the People Programme (Marie Curie Actions) of the European Union's
Seventh Framework Programme FP7/2007-2013/ under REA Grant Agreement No. 317089.
IC acknowledges her student scholarship from the Dinu Patriciu Foundation, Open Horizons, which
supported part of her studies.
During the course of this work, EP was partially supported by the German Research Foundation (DFG) via the Emmy Noether program ``Exact results in Gauge theories''  as well as by the Marie Curie grant FP7-PEOPLE-2010-RG. MT is supported by the RIKEN iTHES project.


\appendix

\section{$SU(2)$ branching points}
\label{app:B}

It is useful to ``order'' the solutions of the discriminant
 \eqref{quadraticdiscriminant}
\be
\Delta =  (t^2\fc^{(k)}_L-u_kt+q\fc^{(k)}_R)^2-4(t-1)(t-q)(t^2\fc^{(2k)}_L+u_{2k}t+q\fc^{(2k)}_R) ~.
\ee
First we take the most extreme case: the massless limit and the $k=1$ case. Which  corresponds to the massless limit of the $\cN=2$ $SU(2)$ with $N_f=4$ where $u_k=0$.
In this limit, the discriminant simplifies drastically to
\be
\Delta =-4 (t-1)(t-q) u_{2k}t ~,
\ee
and its solutions are ordered as
\be
\label{masslessconvention}
t_1 = \infty~, \quad t_2 = 1~, \quad t_3 = q~, \quad t_4 = 0  \, .
\ee
This ordering holds for $q<1$.

For general orbifold $k>1$, but still in the massless  case
\be
\label{discrimSU2}
\Delta = u_k^2 t^2 -4 (t-1)(t-q) u_{2k}t ~ =0
\ee
if we further take the week coupling limit the solutions are ordered as
\be
\label{masslessconvention}
t_1 = \infty~, \quad t_2 =\mathcal{O}(1)~, \quad t_3 = \mathcal{O}(q)~, \quad t_4 = 0  \, .
\ee
with
\be
t_2 = \frac{u_{k}^2+4 u_{2k}}{4 u_{2k}} = \frac{\left(a_1^k+a_2^k\right){}^2}{ a_1^k a_2^k} -4   =  \mathcal{O}(1) \, .
\ee

\bibliographystyle{JHEP}
\providecommand{\href}[2]{#2}\begingroup\raggedright\endgroup

\end{document}